\documentclass[12pt]{article}
\usepackage{graphicx}
\usepackage{color}
\newcommand{\beq}{\begin{equation}}
\newcommand{\eeq}{\end{equation}}
\newcommand{\bea}{\begin{eqnarray}}
\newcommand{\eea}{\end{eqnarray}}

\newcommand{\gsim}{\lower.7ex\hbox{$
\;\stackrel{\textstyle>}{\sim}\;$}}
\newcommand{\lsim}{\lower.7ex\hbox{$
\;\stackrel{\textstyle<}{\sim}\;$}}

\newcommand{\eod}{\end{document}}

\begin{document}
\thispagestyle{empty}
\vspace*{-22mm}

\begin{flushright}
UND-HEP-15-BIG\hspace*{.08em}01\\
Version 9.00 \\


\end{flushright}
\vspace*{1.3mm}

\begin{center}
{\Large {\bf ``Could Charm (\& $\tau$) Transitions be the ``Poor Princess"
Providing a Deeper Understanding of Fundamental Dynamics?"
\\
or:\\
``Finding Novel Forces"}}

\vspace*{10mm}

{Ikaros I.~Bigi}\\
\vspace{5mm}
{\sl Department of Physics, University of Notre Dame du Lac, Notre Dame, IN 46556, USA}\\
{\sl E-mail address: ibigi@nd.edu} \\
\vspace{5mm}
REVIEW ARTICLE, Front. Phys. 10, 101203 (2015) \\
DOI 10.1007/s11467-015-0476-y \\
{\sl Received April 2, 2015; accepted May 4, 2015}

\vspace*{5mm}

\begin{figure}[h!]
\begin{center}
\includegraphics[width=6cm]{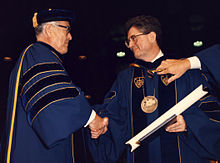}
\end{center}
\caption{The `Team'}
\label{fig:NDTT}
\end{figure}

{\em Dedicated to Timothy O'Meara, mathematician \& first lay Provost of the University of Notre Dame du Lac: without him \& his vision the University of Notre Dame du Lac would be far from what it is today. For example, he opened Notre Dame's academic and cultural doors to China.}

\vspace*{5mm}

\vspace*{10mm}

{\bf Abstract}
\vspace*{-1.5mm}
\\
\end{center}
We know that our Universe is composed of only $\sim$ 4.5\% ``known" matter; therefore, our understanding is incomplete. This can be seen directly in the case of neutrino oscillations (without even considering potential other universes).
Charm quarks have had considerable impact on our understanding of known matter, and quantum chromodynamics (QCD) is the only local quantum field theory to describe strong forces. It is possible to learn
novel lessons concerning strong dynamics by measuring rates around the thresholds of
$[\bar Q Q]$ states with $Q = b, c$. Furthermore, these states provide us with gateways towards
new dynamics (ND),
where we must transition from ``accuracy" to ``precision" eras. Finally, we can make connections with
$\tau$ transitions and, perhaps, with dark matter. Charm dynamics acts as a bridge between the worlds of light- and heavy-flavor hadrons (namely, beauty hadrons), and finding {\em regional asymmetries} in 
many-body final states may prove to be a ``game changer".
There are several different approaches to achieving these goals: for example, experiments such as the 
Super Tau-Charm Factory, Super Beauty Factory, and the Super $Z^0$ Factory act as
gatekeepers -- and deeper thinking regarding symmetries.

\vspace{3mm}

{\bf Keywords;} CKM matrix, HQE \& OPE, CPV in $\Delta C \neq 0 \neq \Delta B$ \& 
$\tau$ decays


\vspace{3mm}

\hrule

\tableofcontents
\vspace{5mm}

\hrule\vspace{5mm}

\section{Landscapes for fundamental dynamics}
\label{History}

At the end of the previous millennium, we realized that
the Universe consists of greater variety than previously believed:
known matter $\sim$ 4.5\%, dark matter $\sim$ 26.5\%, and vacuum (or
dark) energy $\sim$ 69\%. Since the beginning of this millennium,
we have had the following knowledge regarding known matter:

(a) We have {\em failed} to understand the extremely large asymmetry
between known matter and anti-matter in our Universe.

(b) The Standard Model (SM) produces the leading source of the
measured charge parity (CP) violations in neutral kaons and $B$
transitions at least (except, possibly, in $B_s$ oscillations).

(c) No CP asymmetry has yet been established in charm hadron or
baryon decays in general (apart from human existence).

(d) The neutral Higgs-like state has been found in the SM predicted
mass region, and no sign of new dynamics (ND) has been observed in
its decays as of yet. However, we know that the Higgs' amplitude is
primarily a scalar.

(e) It is possible that the impact of Dark Matter {\em may} be
observed, in particular in CP asymmetries in charm hadrons decays. 
Furthermore, $\tau$ lepton decays may be used to calibrate those
correlations. At minimum, we will learn novel lessons about
non-perturbative QCD.

We know that the SM cannot produce neutrino oscillations; this was
found with $\Delta m(\nu_i) \neq 0$ \& three non-zero angles. There
is a reasonable chance of finding CP asymmetries in that case,
despite the background nuclei and anti-nuclei asymmetries. Note
that the definition of known matter is ``fuzzy'' or ``subtle''.

In this review, I primarily focus on measured or measurable charm
hadron transitions, but I do not ignore other phenomena and the
information they can provide. Even when we {\em cannot} establish
the existence of ND in these transitions, we learn novel lessons
about the connections between strong \& weak forces and the dynamics
of beauty hadrons. In other words, the fundamental dynamics around
thresholds of $\bar H_c H_c$ \& $\tau^+\tau^-$ are complex and also
provide indirect information about $H_b$ transitions, where $H_c$
and $H_b$ are heavy mesons containing a heavy ($c$ and $b$) and a
light quark.

First, I will ``paint'' a picture of the flavor dynamics landscape.
Charm quarks have changed the understanding of fundamental dynamics
in several ways.

\begin{itemize}
\item 
Previously, quarks were primarily seen as a mathematical trick to
describe the strong forces between hadrons. Not all researchers
agreed with this concept, however. The charm quark was introduced
for a simple reason, i.e., to describe connections between two quark
and lepton families \cite{BJ}. In 1970, it was suggested as a means of
solving the subtle problem of flavor-changing neutral currents {\em
without} tree diagrams \cite{GIM}. The ``Glashow-Iliopoulos-Maiani
(GIM)'' researchers gave the name ``charm'', meaning to have ``magic
powers'' to prevent bad luck -- like charming a venomous snake
cobra.

\item 
A very good candidate event for the decay of a charm hadron was
found by a group led by Niu in 1971, in emulsion exposed to cosmic
rays and analyzed at Nagoya University \cite{NIU}. $X^{\pm} \to h^{\pm}
\pi^0$ was found, with $h^{\pm}$ denoting a charged hadron that can
be a meson or baryons. With a lifetime of a few $10^{-14}$ s, this
is a weak decay; if $h^{\pm}$ is a meson, the mass of $X^{\pm}$ is
approximately 1.8 GeV. Actually, quarks were already seen as {\em
physical} states by the physics department at Nagoya University;
elsewhere, this concept was mostly ignored.

\item 
In fact, it had already been pointed out in 1963 in the Russian
version of Okun's book \cite{OKUN63}, which was published {\em before} the
discovery of CP violation (CPV), that charm hadrons could be found
in multi-lepton events in neutrino production. Evidence for their
existence was found by interpreting opposite-sign dimuon events:
$\nu N \to \mu^- D... \to \mu^-\mu^+...$ \cite{DIMUONS}.

\item 
In a seminal 1973 paper, Gaillard and Lee \cite{LEE} explored in detail how
charm quarks affect $K^0 - \bar K^0$ oscillations, $K_L \to
\mu^+\mu^-/2\gamma$ through quantum corrections; their findings
yielded a bound of $m_c \leq 2$ GeV. Together with Rosner, they
extended this analysis in a review that was published in the summer
of 1974 \cite{ROSNER74}. At the same time, it was suggested that charm and
anti-charm quarks form an unusually narrow vector meson bound-state,
as a result of gluons carrying three colors and their couplings
decreasing with increasing mass scales \cite{APPLE}. Thus, the theoretical
tools were in place to interpret the surprising observations that
were to come. However, these reports did not convince the skeptics;
they required a Damascus experience to change from ``Saulus'' into
``Paulus'', i.e., from disbelievers into believers.

\item 
Evidence was provided when an unusually narrow resonance in $e^+e^-$
collisions was detected at Stanford Linear Accelerator Center (SLAC)
on the west coast of the US and $p\, Be$ collisions at Brookhaven
National Laboratory (BNL) on the east coast in 1974. This narrow
resonance produced an important ``paradigm'' shift, specifically,
$J/\psi (1S)$ was seen as a boundstate $[\bar cc]$ (after passionate
discussions for a year). This was also established by $\psi (2S)$
and $\psi (3.77)$; the latter produces a $D^0 \bar D^0$ and $D^+D^-$
factory. As previously stated, I refer to this event as the
``October revolution of 1974'' in fundamental dynamics.

\item 
Quarks are real physical states, but they can only be observed in
boundstates, and are not free as named due to
``confinement''\footnote{There is a
subtle exception: top quarks decay before they can produce
boundstates \cite{TOPDECAY}.}. For several reasons, it was realized that
unbroken local color $SU(3)_C$ describes ``strong'' forces from long
to short distances.

\item 
First, it was thought that two pairs of SM quarks were required,
namely, up- \& down-type quarks with $(u,c)$ \& $(d,s)$ with charged
$+2/3$ \& $-1/3$, respectively, named $s=$``strange'' \&
$c=$``charm''.

\item 
On the other hand, the situation at that energy scale was and (\&
still is) considerably more complex, as mentioned above. After many
more discussions \& more careful analyses, researchers realized that
the third lepton family with the charged $\tau$ had already been
found. This also suggested that a third quark family existed, it
``simply'' had to be found.

\item 
The Proceedings of the CCAST Symposium were produced by the
Institute of High Energy Physics (Beijing) in 1987 \cite{YE}. I may
appear to be biased in this regard; however, the Proceedings remain
useful, and not only as regards the history of the field. We have
made sizable progress in the past 27 years, but not in every
respect, and careful readers of these Proceedings still find
directions (or at least signposts) for future progress.

\item 
Wolfenstein introduced the super-weak scenario in 1964
\cite{WOLF64}. This scenario defined the CPV classes, but it is {\em not} a
theory. In retrospect, this means that theorists were slow to deal
with that challenge. Kobayashi \& Maskawa published a paper in 1973
\cite{KM} that discussed the general landscape of CP asymmetries. From
the beginning of the 21st century, we knew that the SM produces the
leading source of the measured CPV at least, with three quark
families (or more). Researchers obtained six triangles with
different shapes, but with the same areas.
\end{itemize}

The book ``A cicerone for the physics of charm'' \cite{CICERONE} relates the
history of high-energy physics regarding flavor dynamics, and also
significantly more: it indicates directions for future research. I
will refer to it several times to aid readers; furthermore, very
interested readers can see peruse the list of references (such as
pioneering papers by Shifman \& Voloshin \cite{SHIFVOL}). Charm hadrons are
primarily seen as somewhat heavy-flavor particles. Charm hadrons
act as the bridge between the worlds of the light- and heavy-flavor
hadrons. This means that the flavor depends on various factors. 
Often, it helps to understand both strong \& weak dynamics. The
research status regarding these topics has changed, as we currently
have significantly more data along with superior analysis tools;
furthermore, theoretical tools have evolved with more focus on
accuracy \& correlations with other techniques.

In the 21st century, one can use models as the first or second steps
to probe data only. More refined theoretical tools have appeared
that are fully based on quantum field theory: operator product
expansion ({\bf OPE}), heavy quark expansion ({\bf HQE}), sum rules
(such as light-cone sum rules), dispersion relations, $1/N_C$
expansions, hybrid renormalization, nonrelativistic QCD (NRQCD),
lattice QCD (LQCD), etc. \cite{CICERONE}. Both judgment and experience are
crucial in determining which tools can be best applied to a given
problem. The SM is not incorrect, but it is obviously incomplete,
and the impact of ND is subtle. The possible dynamics landscape is
``complex''; however, I will focus on items of importance based on
my own judgment:
\begin{itemize}
\item 
The elements $V_{cs}$ and $V_{cd}$ of the  Cabibbo--Kobayashi--Maskawa (CKM) 
matrix must be
accurately measured and connected with other amplitudes. A general
statement can be made: we must focus on precision in $\Delta B  \neq
0$ and accuracy in $\Delta C,\, \Delta S  \neq 0$. First, we must
apply a refined parametrization of the CKM matrix.

\item The meaning of the term ``symmetry'' is broad, but it is taken to
mean CPT (charge conjugation (C), parity transformation (P), time
reversal (T)) invariance here.\ It refers to local symmetries
(unbroken \& broken), global symmetries such as $SU(3)_{\rm fl}$ or
its $SU(2)_{I,U,V}$, discrete symmetries such as P, C, \& CP and
their asymmetries. One can see the difference between local vs.
discrete symmetries in the real world, that is, in physics vs.
chemistry scenarios. Furthermore, one can see the use of
connections between different classes of symmetries in
architecture.\ For example, the Piazza del Campidoglio in the center
of Rome, which was designed by Michelangelo (see Fig.\ref{fig:ROME2}). 
Michelangelo had a subtle and detailed understanding of symmetries
and could manage existing ``backgrounds''. 
\begin{figure}[h!]
\begin{center}
\includegraphics[width=6cm]{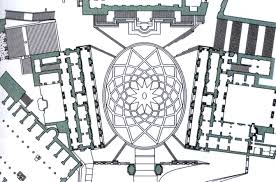}
\end{center}
\caption{Combining different symmetries}
\label{fig:ROME2}
\end{figure}

\item QCD is the only local quantum field theory we have for ``strong''
forces. We must test our control over it {\em quantitatively}, by
examining charm meson \& baryon lifetimes, inclusive semi-leptonic
branching ratios, etc.

\item Measuring $D^+_{(s)} \to \mu^+ \nu (+\gamma's)$, $\tau^+
\nu(+\gamma's)$ provides us with superior tests of LQCD and also,
perhaps, evidence for ND.\ For exclusive semi-leptonic decays, the
scenarios are more complex, since long-distance dynamics are
crucial. LQCD and other theoretical tools are furnished with very
good test grounds.

\item Very suppressed decays such as $D^0 \to \gamma \gamma$, $D_{(s)} \to
l^+l^- X$, etc., have not been found to date. This issue is
dominated by long-distance dynamics over which we have little
control. If sufficient large datasets were provided, we might learn
from those rates. However, when we have to analyze refined
asymmetries, we have an opportunity to determine the existence of ND
\cite{PPB}.

\item It is important to find CP asymmetries in two-body final states (FS)
in mesons \& baryons. However, it is crucial to probe {\em
regional} CP asymmetries in three- \& four-body FS. We have two
examples with $B^{\pm}$ decays \cite{LHCb028,IRINA}. The SM produces very small
CP asymmetries in singly Cabibbo-suppressed (SCS) transitions and
basically zero in doubly Cabibbo-suppressed (DCS) ones. In the
latter we could find impact of ND with hardly one SM background.
\end{itemize}
The usual tools for strong-force investigations provide a good
spectroscope for hadrons. However, when we include weak dynamics,
more refined tools are required in order to probe CP asymmetries.

This article is organized as follows: in Section 2, I discuss local
and global symmetries and the tools required for heavy hadron data
analyses in general; I discuss (semi-)leptons and rare decays of
charm hadrons in Section 3; then, in Section 4 I turn to the
non-leptonic decays of charm hadrons. These provide a significantly
more complex landscape, in particular regarding many-body FS. It is
important to calibrate $\tau$ decays in $\Delta S=0,1$ for several
reasons, as shown in Section 5. Comments about correlations with
beauty transitions are given in Section 6. Finally, Section 7
summarizes the current status of the field and provides an outlook
for the future.

\section{Symmetries \& tools}
\label{TOOLS}

The flavor landscapes differ significantly for charm \& beauty hadrons and $\tau$
leptons with different uncertainties. Obviously,
beauty hadrons carry heavy flavor; however, charm hadrons and $\tau$ lepton mostly are on the 
right side of heavy flavor. 
In my view, there is a more general term: ``symmetry"
(= ``$\sigma \upsilon \mu \mu \epsilon \tau \rho \alpha$") goes beyond the meaning of ``tools".

\subsection{Refined parametrization of CKM matrix}
\label{CKM}

The dynamics of flavor violation in the SM world are
described by the CKM matrix as the first step. The CKM matrix
describes the quark couplings with left-handed charged bosons in
terms of three angles and one weak phase with three families. The
correlations between the matrix elements are described by three
unity relations and six triangles, where the latter have the same
area \footnote{A general statement
regarding the numbers $N$ of quark families can be made. For $N=2$,
there is no CPV source. For $N=4$ (or more), the landscape is 
significantly more
``complex'' and triangles are insufficient. In other words, one
probes triangles regardless of whether the
sum of their angles is $180^\circ$.}. We primarily focus on hadron
decays, but the CKM matrix is also used for the productions of
flavor hadrons with
\bea
\sum_{i=1}^{3} |V_{ij}|^2 = 1 \; \; &;& \; \; j=1,2,3, \\
\sum_{i=1}^{3}V_{ji}V^{*}_{ki}   = 0 = \sum_{i=1}^{3}V_{ij}V^{*}_{ik}
\; &;& \; j,k = 1,2,3,  \; j\neq k, \\
|{\rm Im}V^*_{km}V_{lm}V_{kn}V^*_{ln}| =
 |{\rm Im}V^*_{mk}V_{ml}V_{nk}V^*_{nl}|  = J &,&
{\rm area\, (every\, triangle)} = \frac{1}{2} J.
\eea
In the SM with three quark families, these equalities are correct
(excluding experimental uncertainties). 
The majority of researchers use the ``smart"
parametrization going back to Wolfenstein \cite{WOLFMAT} with an obvious pattern although we did not 
understand its source. 
This approach involves three parameters: $A$, $\bar \rho$, and $\bar \eta$, which are assumed to be of order unity, and a known $\lambda = {\rm sin}\theta_C \sim 0.22$, to be used for expansions in higher orders. This approach describes the flavor dynamics data quite well, including CP violation. This parametrization puts six triangles into three classes and is very successful: 
\begin{eqnarray}
{\bf {\rm V}}_{\rm CKM} \simeq
\left(\footnotesize
\begin{array}{ccc}
 1 - \frac{\lambda ^2}{2}  & \lambda , &
 A\lambda^3(\rho - i\eta +\frac{i}{2}\eta \lambda ^2)  \\
-\lambda & 1 - \frac{\lambda ^2}{2} - i\eta A^2\lambda^4 &
A\lambda^2(1+i\eta\lambda^2)
\\
A\lambda^3(1-\rho - i\eta) & -A\lambda^2 &1
\end{array}
\right),
\end{eqnarray}
\bea
{\rm `Old' \, triangle\; I.1:}&&V_{ud}V^*_{us} \; \; \;  [{\cal O}(\lambda )] + V_{cd}V^*_{cs} \;  \; \;
[{\cal O}(\lambda )] +
 V_{td}V^*_{ts} \; \; \; [{\cal O}(\lambda ^{5} )] = 0,   \\
{\rm `Old' \, triangle\; I.2:}&& V^*_{ud}V_{cd} \; \; \;  [{\cal O}(\lambda )] + V^*_{us}V_{cs} \; \; \;  
[{\cal O}(\lambda )] +
V^*_{ub}V^*_{cb} \; \; \; [{\cal O}(\lambda ^{5} )] = 0,    \\
{\rm `Old' \, triangle\; II.1:}&& V_{us}V^*_{ub} \; \; \;  [{\cal O}(\lambda ^4)] + V_{cs}V^*_{cb} \;  \; \;  
[{\cal O}(\lambda ^{2} )] +
V_{ts}V^*_{tb} \; \; \; [{\cal O}(\lambda ^2  )] = 0,   \\
{\rm `Old' \, triangle\; II.2:}&& V^*_{cd}V_{td} \; \; \;  [{\cal O}(\lambda ^4 )] + V^*_{cs}V_{ts} \; \; \;  
[{\cal O}(\lambda ^{2})] +
V^*_{cb}V^*_{tb} \; \; \; [{\cal O}(\lambda ^{2} )] = 0,  \\
{\rm `Old'\, triangle\; III.1:}&& V_{ud}V^*_{ub} \; \; \;  [{\cal O}(\lambda ^3)] + V_{cd}V^*_{cb} \;  \; \;  
[{\cal O}(\lambda ^{3} )] +
V_{td}V^*_{tb} \; \; \; [{\cal O}(\lambda ^3  )] = 0,   \\
{\rm `Old' \, triangle\; III.2:}&& V^*_{ud}V_{td} \; \; \;  [{\cal O}(\lambda ^3 )] + V^*_{us}V_{ts} \; \; \;
[{\cal O}(\lambda ^{3})] +
V^*_{ub}V^*_{tb} \; \; \; [{\cal O}(\lambda ^3 )] = 0.
\label{NEW2}
\eea
Fitting global 2014 data and using $\bar \rho = \rho (1 - \lambda ^2 /2 + ... )$, etc., gives \cite{PDG14}
\bea
\lambda = 0.22537 \pm 0.00061 \; \; \;  &,& \; \; \;    A = 0.814 ^{+0.023}_{-0.024}, \\
\bar \rho = 0.117 \pm 0.021 \; \; \; &,& \; \; \; \bar \eta = 0.353 \pm 0.013.
\label{2014}
\eea
However, one subtle problem occurs. The data suggests that $|\bar \eta|$ and $|\bar \rho |$ are 
{\em not} of order unity, with the latter being further removed. It is somewhat surprising how
this obvious pattern is so successful, despite its disagreement with the expected values
of $\bar \eta$ and $\bar \rho$. In the present era, accuracy and even precision are required.
Other parametrizations have therefore been suggested and for good reasons. For example, the method proposed in
Ref.\cite{AHN} uses $\lambda$ with
$f \sim 0.75$, $\bar h \sim 1.35$, and $\delta_{\rm QM} \sim 90^o$.
This approach is close to reality as regards incorporating a non-leading source for $B$ decays and/or a very small source for $D$ decays in the SM, with
\begin{eqnarray}
V_{\rm refined}= \left(\footnotesize \begin{array}{ccc}
V_{ud} & V_{us} & V_{ub} \\
V_{cd} & V_{cs} & V_{cb} \\
V_{td} & V_{ts} & V_{tb}
\end{array}\right)     =
\; \; \; \; \;  \; \; \; \; \;  \; \; \; \; \; \; \; \; \; \;  \; \; \; \; \; \; \; \; \; \; \; \; \; \; \; \;
&&   \\
=\left(\footnotesize
\begin{array}{ccc}
 1 - \frac{\lambda ^2}{2} - \frac{\lambda ^4}{8} - \frac{\lambda ^6}{16}, & \lambda , &
 \bar h\lambda ^4 e^{-i\delta_{\rm QM}} , \\
 &&\\
 - \lambda + \frac{\lambda ^5}{2} f^2,  &
 1 - \frac{\lambda ^2}{2}- \frac{\lambda ^4}{8}(1+ 4f^2)
 -f \bar h \lambda^5e^{i\delta_{\rm QM}}  &
   f \lambda ^2 +  \bar h\lambda ^3 e^{-i\delta_{\rm QM}}   \\
    & +\frac{\lambda^6}{16}(4f^2 - 4 \bar h^2 -1  ) ,& -  \frac{\lambda ^5}{2} \bar h e^{-i\delta_{\rm QM}}, \\
    &&\\
 f \lambda ^3 ,&
 -f \lambda ^2 -  \bar h\lambda ^3 e^{i\delta_{\rm QM}}  &
 1 - \frac{\lambda ^4}{2} f^2 -f \bar h\lambda ^5 e^{-i\delta_{\rm QM}}  \\
 & +  \frac{\lambda ^4}{2} f + \frac{\lambda ^6}{8} f  ,
  &  -  \frac{\lambda ^6}{2}\bar h^2  \\
\end{array}
\right)
+ {\cal O}(\lambda ^7).
\label{MATRIX}
\end{eqnarray}
Thus, the landscape of the CKM matrix is more subtle than usually stated; it is described by six triangles that differ subtly in ways, but retain the same area. Therefore,
\bea
{\rm Triangle\; I.1:}&&V_{ud}V^*_{us} \; \; \;  [{\cal O}(\lambda )] + V_{cd}V^*_{cs} \;  \; \;  [{\cal O}(\lambda )] +
 V_{td}V^*_{ts} \; \; \; [{\cal O}(\lambda ^{5\& 6} )] = 0,   \\
{\rm Triangle\; I.2:}&& V^*_{ud}V_{cd} \; \; \;  [{\cal O}(\lambda )] + V^*_{us}V_{cs} \; \; \;  [{\cal O}(\lambda )] +
V^*_{ub}V^*_{cb} \; \; \; [{\cal O}(\lambda ^{6 \& 7} )] = 0,    \\
{\rm Triangle\; II.1:}&& V_{us}V^*_{ub} \; \; \;  [{\cal O}(\lambda ^5)] + V_{cs}V^*_{cb} \;  \; \;  [{\cal O}(\lambda ^{2 \& 3} )] +
V_{ts}V^*_{tb} \; \; \; [{\cal O}(\lambda ^2  )] = 0,   \\
{\rm Triangle\; II.2:}&& V^*_{cd}V_{td} \; \; \;  [{\cal O}(\lambda ^4 )] + V^*_{cs}V_{ts} \; \; \;  [{\cal O}(\lambda ^{2\& 3})] +
V^*_{cb}V^*_{tb} \; \; \; [{\cal O}(\lambda ^{2 \& 3} )] = 0,  \\
{\rm Triangle\; III.1:}&& V_{ud}V^*_{ub} \; \; \;  [{\cal O}(\lambda ^4)] + V_{cd}V^*_{cb} \;  \; \;  [{\cal O}(\lambda ^{3\& 4} )] +
V_{td}V^*_{tb} \; \; \; [{\cal O}(\lambda ^3  )] = 0,   \\
{\rm Triangle\; III.2:}&& V^*_{ud}V_{td} \; \; \;  [{\cal O}(\lambda ^3 )] + V^*_{us}V_{ts} \; \; \;
[{\cal O}(\lambda ^{3\& 4})] +
V^*_{ub}V^*_{tb} \; \; \; [{\cal O}(\lambda ^4 )] = 0.
\label{NEW2}
\eea
The pattern in flavor dynamics is less obvious for CP violation in hadron decays, as stated previously \cite{BUZIOZ}. Triangles III.1, II.1, and I.1 describe $B^0$, $B^0_s$, and $K^0$, respectively, including
oscillations.
Super-heavy top quarks decay before they can produce hadrons \cite{TOPDECAY}, and Triangle I.2 affects charm transitions. SCS transitions provide a more complex scenario, as can be seen in $c \to d u \bar d$, $c\to s u \bar s$ diagrams and $c \Rightarrow u$ transitions. The latter poses a veritable challenge as regards connecting quark diagrams with hadronic amplitudes; for example, the difference between penguin diagrams and final state interactions ({\bf FSI})/re-scattering is ``fuzzy". Furthermore, we must consider
interference between Cabibbo-favored \& DCS amplitudes.

The correlations
between triangles are very important; for example, $c \to d u \bar s $ describes DCS amplitudes
in mesons \& baryons and gives zero weak phases up to ${\cal O}(\lambda ^7)$. I will discuss this and
connections with beauty hadron transitions below. 
This is only the first step in discussing the information that the data give us. 
A second step is also required, which involves both work and judgment. A third step is necessary, in which additional data, tools, time, and thinking are required.

\subsection{Adler-Bell-Jackiw (ABJ) (or triangle) anomaly}
\label{ABJ}

In the world of three quark and two lepton families, another subtle challenge had to be overcome. A classical symmetry is expressed because of the existence of a conserved
current; we obtain $\partial ^{\mu}J^5_{\mu} = 0$ for massless fermions.
However, the triangle diagram with an internal loop of only fermions coupled to
three external axial vectors or one axial \& two vectors generates a ``quantum anomaly"; i.e.,
it {\em removes} a classical symmetry \cite{ANOM}:
\beq
\partial ^{\mu}J^5_{\mu} = \frac{g_S^2}{16\pi^2} G\cdot \tilde G \neq 0,
\eeq 
even for {\em massless} fermions. $G$ and $\tilde G$ denote the gluonic field strength tensor
\& its dual, where $\tilde G_{\mu \nu} = \frac{i}{2}\epsilon _{\mu \nu \rho \sigma}G^{\rho \sigma}$.
$G\cdot \tilde G \neq 0$ by itself yields a finite result, yet it destroys the renormalizability
of the theory. That is, it cannot be ``renormalized away" in a gauge-invariant manner with a dimensional four
operator. Instead, it must be neutralized by adding a contribution from all fermion classes
in the theory to obtain a zero result. For the SM, the sum of all electric charges of fermions of a given
family must be zero. This imposes a connection between the quark \& lepton charges, i.e.,
$e$ \& $\mu$ have charge number ``-2", and the
$u,d,s$ quarks with three colors ``0"; however, with colored {\em charm} quarks we obtain ``+2".
This result is excellent, yet the connection is unexplained. Another challenge exists, as we have found that the $\tau$ lepton adds another
charge number ``-1" with a mass similar to those of charm mesons. Therefore, some researchers expected
to find the third family of quarks, namely, $[t,b]$, with significantly heavier
masses. This indicates that Nature has a sense of humor to deal with our understanding or lack of it.

There are three points to note here: (a) The impact of the ``ABJ anomaly" has an unusually long history in modern physics: these important papers were published over 45 years ago \cite{ANOM}. 
(b) This anomaly did not only have theoretical implications. It also had an impact in the real world, in relation to the
$\pi^0 \to 2 \gamma$ decay in particular. In fact, an even older paper by
Nobel Prize Winner J. Steinberger \cite{STEIN} discusses this point. (c) The ``ABJ anomaly" is not primarily relevant in terms of history; one learns from the
theoretical techniques used previously and applied in other landscapes.

\subsection{Theoretical tools for decays}
\label{THTOOL}

I assume  CPT invariance, analyticity, \& unitarity in quantum field theory (\& also in effective theories).  
These connections are subtle
in many ways. There are three classes of
FS for hadrons: leptonic, semi-leptonic, and non-leptonic
\footnote{The first class applies to mesons only.}. Furthermore, there are both inclusive and exclusive subclasses, where different tools (with different uncertainties) can
be used. I will return to this topic below and discuss it in some detail. The same classifications apply
to the decays of both charm \& beauty hadrons, although the details differ;
for example, Dalitz plots for three-body FS are primarily populated with charm decays, while the
center is basically empty of beauty hadrons. Finally, one can and should use semi-hadronic
$\tau$ decays to calibrate predictions with real data.

Quark diagrams are described with two-dimensional plots; however, in general, the FS
are described by three-dimensional plots (and beyond, when one includes spin observables).
To be realistic, it is sufficient to discuss nonleptonic decays with four-body FS at most.
Furthermore, the connections between quark diagrams and operators are subtle, particularly as regards local and non-local operators. Note that the latter depend crucially on long-distance {\bf FSI}. 
I will discuss these classes in more detail below. 
\begin{itemize}
\item
First, one focuses on two-body non-leptonic FS. 
These states give one-dimensional observables from the rates and numbers of CP asymmetries.
\item
Probing Dalitz plots for CP asymmetries gives two-dimensional observables, as we have previously seen
regarding $B$ decays.   
I will comment on this below.
If the plot is not flat, it indicates that the {\bf FSI} are not trivial, and are similar to resonances in different ways.
One applies amplitudes for FS with hadrons and resonances, with
$P \to h_1[h_2h_3] + h_2[h_1h_3] + h_3[h_1h_2] \Rightarrow h_1h_2h_3$
\footnote{Fans of ballet know that ``pas de deux" is an important dance and must be performed by experts, but one also requires ``pas de trois" and ``pas de quatre", as for charm dynamics.}. I am not claiming
that three-body amplitudes are perfectly described by a sum of two-body FS. However, this approach is sufficient to a large extent, realistically speaking. 

As a second step 
that the analyses are model-insensitive \footnote{Subtle differences exist between
``insensitive" and ``independent", as discussed previously.}. 
However, we must remember that the real theory does not always yield the best fitting of the data. Furthermore, we must measure correlations with other data. 
We have the tools to measure regional asymmetries in Dalitz plots. First, one uses model-insensitive tools, and then real theoretical tools that are validated
based on correlations with other transitions are applied. Thus, these theoretical tools must be ``acceptable". Note that the criteria determining acceptability vary.
\item
One must be realistic with finite data when probing four-body FS and identify tools to analyze
one-dimensional asymmetries. We are at the beginning of the road towards understanding the
underlying forces.
\end{itemize}
The real impact of ND will become apparent in detailed discussion.

Connections between effective quark operators and hadronic transitions due to
``duality" exist \cite{DUAL} -- however, they are subtle. One cannot
compare the FS using measured hadron masses and suggested mass values for quarks only; this neglects the crucial point of duality, i.e., the impact of {\em non}-perturbative forces.

The landscapes of CP asymmetries in charm (\& beauty) hadrons provide a ``wonderful challenge" for probing ND (including baryon decays \cite{BARIB}). At minimum, we learn about the impact of 
{\bf FSI} in the world of hadrons.

For several reasons, the number of colors must be three (specifically, neither two nor four). Yet, in the
limit of $N_C \to \infty$, QCD's non-perturbative dynamics becomes tractable \cite{HOOFT}. Thus, only planar diagrams contribute to hadronic scattering, and the asymptotic states are $\bar qq=$ mesons
\& $qqq=$ baryons. ``Confinement" is then proven (also $\bar q q \bar q q$, etc.). Further, the Zweig or OZI rule holds. One treats short-distance dynamics with $N_C = 3$
fixed, so as to derive an effective Lagrangian at lower scales. Once the Lagrangian has been devolved
to the scale at which one wishes to evaluate the hadronic matrix elements, which are shaped by
long-distance dynamics, one expands the matrix elements in powers of $1/N_C$ for $H_Q \to f$, such that
\beq
\langle f |{\cal L}_{\rm eff}|H_Q\rangle \propto b_0 + \frac{b_1}{N_C} + {\cal O}(1/N_C^2).
\eeq
This expansion of $N_C \to \infty$ has often indicated the aforementioned directions for future research. For example, it has aided researchers in treating two-body non-leptonic decays of charm mesons \cite{CICERONE,IBCHINA}.
This technology lies between models where one can discuss uncertainties {\em inside} the model and real theories, where
the uncertainties can be decreased systematically. It is not truly an expansion, since it cannot go
{\em beyond} $b_1$.

\subsubsection{Effective transition amplitudes {\em including} re-scattering}
\label{EFFECT}

One can describe the amplitudes of hadrons
with CPT invariance following the history outlined above; it is given
in detail in Refs.\cite{1988BOOK,WOLFFSI} and
in Sect. 4.10 of Ref.\cite{CPBOOK}.
\bea
T(P \to f) &=& e^{i\delta_f} \left[ T_f +
\sum_{f \neq a_j}T_{a_j}iT^{\rm resc}_{a_jf}  \right],
\label{CPTAMP1}
\\
T(\bar P \to \bar f) &=& e^{i\delta_f} \left[ T^*_f +
\sum_{f \neq a_j}T^*_{a_j}iT^{\rm resc}_{a_jf}  \right]   \; ,
\label{CPTAMP2}
\eea
where $T^{\rm resc}_{a_jf}$ describe the {\bf FSI} between $f$ and intermediate
{\em on}-shell states $a_j$ that connect with this FS. It is generally sufficient to focus
on strong re-scattering; one can label it simply {\bf FSI}.
One obtains ``regional" CP asymmetries and not only ``averaged" results, with
\beq
\Delta \gamma (f) = |T(\bar P \to \bar f)|^2 - |T(P \to f)|^2 =
4 \sum_{f \neq a_j} T^{\rm resc}_{a_jf} \, {\rm Im} T^*_f T_{a_j} \; .
\label{REGCPV}
\eeq
CP asymmetries must vanish upon summing over all such $f$ states
using CPT invariance between {\em subclasses} of partial widths, where
\beq
\sum_{f} \Delta \gamma (f) =
4 \sum_{f}\sum_{f \neq a_j} T^{\rm resc}_{a_jf} \, {\rm Im} T_f^* T_{a_j} = 0 \; ,
\eeq
since $T^{\rm resc}_{a_jf}$ \&  Im$T_f^* T_{a_j}$ are symmetric \& antisymmetric,
respectively, in the indices $f$ \& $a_j$. 

These FS $f$ consist of two-, three-, four-body states, etc., such as pions and kaons. One describes
three-body FS using Dalitz plots, whereas the landscapes of four-body states, etc., are even more ``complex", being essentially a ``drama with more actors".
In principle, one can probe local asymmetries, however, one must be realistic regarding finite data and
a lack of ``perfect" quantitative control of non-perturbative QCD. The first step is to use models for looking  
at the data; the second step is to analyze model-insensitive ways. Finally we should 
{\em not} be ``slaves" of the best fits of the data. Instead, we require real theories providing understanding
of the underlying dynamics. We must also consider the correlations between our obtained data and interpret them in an acceptable manner. This statement is subtle (and also concerns the definition of ``regional" asymmetries), but crucial. I will discuss these points in some detail below.

We can describe transitions of boundstates of $\bar q q$ (or $qqq$); the simplest case is for mesons,
but it is not simple. We must include re-scattering due to strong forces 
\footnote{For practical reasons, we can generally ignore quantum electrodynamics (QED) FSI.} and its large impact.
Penguin diagrams can account for absorption due to internal $c$ quarks {\em in principle}
by adding pairs of $\bar qq$ for beauty hadrons. However the cases involving charm hadrons
are unclear, even in principle.
The connections of penguin and tree diagrams with reality are often fuzzy, as pointed out in
Refs. \cite{1988BOOK,WOLFFSI,CPBOOK}.

Can we quantitatively connect quark diagrams with hadronic
amplitudes? It is one thing to draw quark diagrams by adding pairs of
$\bar qq$, but trusting them is a completely separate issue. How can one connect data
concerning decays for two-, three-, four-body FS with information about the
underlying dynamics? We must apply several theoretical tools in this case, which must be connected with other 
transitions, and we must also consider their limits. Here, I will discuss U-spin symmetry, focusing on its uncertainties and its connections with the V-spin case. I will also comment briefly on dispersion relations.

\begin{figure}[h!]
\begin{center}
\includegraphics[width=4cm]{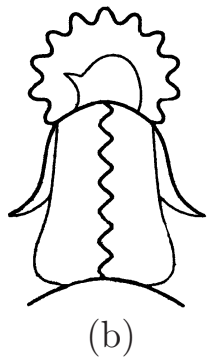}
\end{center}
\caption{Artistic diagrams of penguin amplitudes; the picture of the (b) diagram was reproduced
from {\em Parity} by permission of T. Muta \& T. Morozumi}
\label{fig:nicepeng}
\end{figure}

Penguin diagrams show amplitudes for $Q \to q +$ gluons, where
$Q$ and $q$ quarks carry the {\em same} charge; consider the artistic version shown in Fig.\ref{fig:nicepeng}
with large solid quark lines and wavy lines for $W^{\pm}$ plus one gluon.
Ignoring artistic ambition, penguin diagrams are satisfactory as regards $b$ (\& $s$) quarks. However,
one should not hide theoretical uncertainties; furthermore,
different scenarios exist: $b \Rightarrow s,\, d $ amplitudes
are given by local or short-distance operators and have a sizable impact on inclusive rates. For
exclusive rates, however, we have less control. On the other hand, we have
$c\Longrightarrow u$ amplitudes, which are mostly dominated by long-distance dynamics, where we have less control over inclusive rates and significantly less control over exclusive rates.
Based on chiral symmetries, one expects them to primarily affect two-body FS and
to have some influence on three-body FS, but hardly beyond. Re-scattering amplitudes include the impact of penguin diagrams, but their landscapes are significantly broader:
\begin{itemize}
\item
Eqs. (\ref{CPTAMP1},\ref{CPTAMP2}) apply to amplitudes in general, including many-body FS, whether for hadron or quark boundstates (with constitute quarks) in initial states
(\& between). This is regardless of whether or not we can perform this calculation.

\item
The manner in which one can connect the hadronic and quark amplitude landscapes depends on various factors.
One hopes to be sufficiently removed from the $\bar c c$ threshold to
produce $\Delta \Gamma (B_{s,d})$ for $B_{s,d}$ primarily through short-distance dynamics;
this has also been somewhat suggested for $\Delta \Gamma (D^0)$, perhaps.
When one discusses direct CP asymmetries, one requires both weak \& strong phases.
Quark amplitudes give weak phases, while penguin diagrams from non-local operators provide the
imaginary component that one requires for (strong) re-scattering. However, SCS transitions of charm hadrons are very complex. There is a difference between diagrams
one can compute and amplitudes that are measurable because of interference including
re-scattering.

\item
A general statement can be made. Since our control of strong dynamics is quite limited quantitatively (at present), ``global" strong phases are very often
used to discuss data concerning three- \& four-body FS. It is claimed that accurate information can be obtained in this manner. However, this is only the first step.

\item
Penguin diagrams do {\em not} affect DCS decays of  $D_{(s)}$ \& $\Lambda_c^+$,
while re-scattering {\em does}.

\item
We require the aid of refined tools like dispersion relations to understand the information provided by the data. I will discuss these items below.

\end{itemize}
We must consider which theoretical tools we can apply and their limits.
Obviously, chiral symmetry is an excellent candidate, although some subtle points must be considered.
U-spin symmetry is a ``popular" candidate. However, I have grave concerns regarding
the control of theoretical uncertainties, in particular by ignoring the connections between
U- \& V-spin symmetries and, even worse, FS with {\em only charged} hadrons. I will discuss
this problem below.

\subsubsection{Connections of U- \& V-spin symmetries: spectroscopy vs. weak decays}
\label{USPIN}

The global (\& broken) $SU(3)_{fl}$
with its three subsymmetries $SU(2)_I$, $SU(2)_U$, \& $SU(2)_V$ was introduced first, when ``constituent" quarks were primarily seen as mathematical tools to describe the hadron spectroscopies rather than real physical states.  They are applicable to spectroscopy and can be used to discuss baryon and meson masses, although the latter
are significantly affected  by chiral symmetry. When one compares the masses of nucleons, $\Lambda$ and $\Xi$, one can suggest the values of the constituent quark masses, where
$m_u^{\rm const} \simeq m_d^{\rm const} \sim 0.3$ GeV \& $m_s^{\rm const} \sim 0.5$ GeV
\cite{LIPKINBOOK}. Now, we can compare
the masses of charm baryons; i.e., $M(\Lambda _c^+) \simeq 2.29$ GeV vs.
$M(\Xi_c^+) \simeq 2.46$ GeV and $M(\Xi_c^0) \simeq 2.47$ GeV vs. $M(\Omega_c^0) \simeq 2.7$ GeV.
One obtains differences of $\sim 0.2$ GeV in both cases; therefore, this approach is satisfactory, but this is not an accurate tool. We have a better
understanding of this: the mixing of
$\langle 0| \bar u u|0\rangle$, $\langle 0|\bar d d|0\rangle$
between $\langle 0|\bar s s |0\rangle$ with scalar resonances are {\em not} OZI suppressed \cite{DR}.
It makes sense to use U-spin symmetry when considering the spectroscopy of charm \& beauty hadrons.
However, these situations
are more complex when one combines strong \& weak dynamics.

Re-scattering has an important impact on weak amplitudes in general and on CP asymmetries in particular (see Eqs. (\ref{CPTAMP1} - \ref{REGCPV})) \cite{1988BOOK,WOLFFSI,CPBOOK}.
We cannot ignore the correlations of U-spin with V-spin symmetries. 
In other words, one cannot focus on two-body FS or even more with only charged particles in weak transitions.
Simple situations appear in very low-energy collisions of
$K^-\pi^+ \Leftrightarrow \bar K^0\pi^0$ using $SU(2)_I$ symmetry and even
$K^-\pi^+ \Leftrightarrow \bar K^0\eta$.
However, at somewhat higher energies one must discuss re-scattering, primarily regarding
$K \pi \to K 2\pi,\, K 3\pi $, and even $\pi K \to 3 K, \, 3K\pi$, etc., where obvious differences between the {\em initial} and {\em final} states exist.
This also changes $\pi^+\pi^- \Leftrightarrow \pi^0\pi^0$ at very low energies. However, the
situation changes significantly at slightly higher energies, with
$\pi^+\pi^-/\pi^0\pi^0/\pi^+\pi^0 \to 4 \pi ...$ because of G-parity.
Furthermore, this affects $\pi \pi \Leftrightarrow \bar K K$ at very low energies, but the landscape
is also $\pi \pi \to K \bar K\pi , K\bar K 2\pi,K \bar K K \bar K$ at somewhat higher energies.

There are very different time scales for weak vs. strong forces.
Therefore strong re-scattering has a large impact; it makes the differences
between U- \& V-spin symmetries very fuzzy.
Obviously, U-spin symmetry is broken significantly. The first guess is
$(M_K^2 - M^2_{\pi} ) < (M_K^2 + M^2_{\pi} )$, and more refined solutions are based on the constituent quarks. One can use this approach for models to predict {\em exclusive} decays, but with large theoretical uncertainties; the problem lies in treating the
{\bf FSI} {\em quantitatively}. In particular, we have the tools to probe Dalitz plots
with like dispersion relations. The only problems we must face are the requirements for more data and more
time to analyze these findings and to check them against correlations with other transitions. I will return to this topic and discuss it in some detail below.

In the world of quarks, one describes primarily {\em inclusive} transitions. ``Current"
quarks with $m_u < m_d << m_s$ are based on theory. I-, U- , \& V-spin symmetries consider $u \leftrightarrow d$, $d \leftrightarrow s$, \& $u \leftrightarrow s$, respectively.
These three symmetries are obviously broken on different levels, and these violations are
connected in the SM. The operators producing inclusive FS depend on their CKM
parameters and the current quark masses involved there. However, the real scale for inclusive
decays is given by
the impact of QCD, i.e., $\bar \Lambda \sim  1$ GeV 
\footnote{For good reasons, one uses different and smaller $\Lambda_{\rm QCD} \sim 0.1-0.3$ GeV to describe jets in collisions.}.
Thus, the violations of U- \& V-spin symmetries are small, and tiny for the I-spin case.
We can deal with inclusive rates of beauty and, perhaps, charm hadrons using
effective operators in the world of quarks.

The connections between inclusive and exclusive hadronic rates are not obvious, particularly as regards quantitative techniques. The violations of
I-, U-, \& V-spin symmetries in the measurable world of hadrons are expected to be scaled by the differences in pion and kaon masses,
which are {\em not} small compared to $\bar \Lambda$ (or
$[m^2_K - m^2_{\pi}]/[m^2_K + m^2_{\pi}]$).
This is even more crucial in terms of direct CP violation and the impact of {\em strong} re-scattering on amplitudes.

Returning to the history of this field, Lipkin suggested that U-spin violations in $B$ decays
are of the order of 10--20 \% \cite{LIPK2} in CKM-favored cases, and may be larger in
suppressed cases. One reason for this is that suppressed decays in the world of hadrons consist of larger numbers of states
in the FS, where strong {\bf FSI} with opposite signs have significant impact.
Furthermore, the worlds of
hadrons (or constitute quarks) are controlled by {\bf FSI} because of {\em non}-perturbative QCD; they have the strongest impact on exclusive cases.
For good reasons, it has been stated that violation of U-spin symmetry is
approximately ${\cal O}(10 \%)$ in inclusive decays. In the sum of exclusive decays, large ratios that fluctuate more significantly can be seen, and I will discuss well-known examples of this below.
My central point is that we cannot discuss U-spin symmetry (\& its violations) singly; instead, we must discuss
connections with V-spin symmetry.

\subsection{Expansions}

Usually, we cannot truly solve the challenges we face in the QFT landscapes. Many of the best
theoretical tools we have are based on certain expansions, where some systematic uncertainties exist
\footnote{Of course, they can still be incorrect.}. I am not saying that we cannot use models; however, this is
the first step being taken in the 21st century and the research direction should be changed, based on improved data and more careful thinking. Models have no systematic limits.

I will mention a special case, namely, QCD. First, there is no competition from any other local gauge
theory. It is not trivial at all to combine truly strong forces in long distances with asymptotic freedom
at short distances using this approach. Further, QCD is crucial to combine self-interactions of three and four gluons with their color
quarks, and it is much easier to draw diagrams with gluon-quark couplings. However, one then overlooks the crucial
point of non-abelian gauge theories.

\subsubsection{Heavy quark theory}
\label{HQS}

The lack of full calculational control of strong forces limits our understanding of the information
given by the data. We require other tools, such as chiral theory, to consider non-perturbative dynamics in
special settings. We have heavy-quark symmetry ({\bf HQS}).
The non-relativistic dynamics of a spin-$\frac{1}{2}$ particle with charge $g$ is described by the
Pauli Hamiltonian
\beq
{\cal H}_{\rm Pauli} = - g A_0 + \frac{(i\vec\partial -g\vec A)^2}{2m} +\frac{g\vec \sigma \cdot \vec B}{2m},
\eeq
where $A_0$ \& $\vec A$ denote the scalar \& vector potentials and the magnetic field is $\vec B$. In the
heavy  mass limit, only the first term survives, such that
\beq
{\cal H}_{\rm Pauli} \to - g A_0 \; \;  {\rm as} \; \; m \to \infty \;,
\eeq
i.e., an infinite heavy ``electron" is static. It does not propagate, instead it interacts only via the
``Coulomb" potential and its spin dynamics become decoupled. 

This is also the case for an infinite heavy quark.
Its mass is separate from its dynamics (although not its kinematics), and it is the source of a static color
Coulomb field that is independent of the heavy-quark spin. That is the statement made by the {\bf HQS}. There are several direct consequences of the heavy-light system spectrum, i.e.,
mesons = $[Q\bar q]$ and baryons = $[Qq_1q_2]$. First, in the limit of $m_Q \to \infty$, the spin
of the heavy quark $Q$ decouples, and the spectra of the heavy-flavor hadrons are described in
terms of the spin and orbital degrees of freedom of the {\em light} quarks alone. Therefore, to leading
order accuracy, one obtains no hyperfine splitting
\footnote{In the world of mesons, one can consider comparing the squares of the meson masses, where
$M_{B^*}^2 - M_B^2 \sim 0.49$ (GeV)$^2$ and $M_{D^*}^2 - M_D^2 \sim 0.55$ (GeV)$^2$. However, hyperfine
splittings are somewhat ``universal", i.e., $M_{\rho}^2 - M_{\pi}^2 \sim 0.43$ (GeV)$^2$ and
$M_{K^*}^2 - M_K^2 \sim 0.41$ (GeV)$^2$. Is this simply a fortunate connection between light and heavy mesons,
or have we neglected something?} and
\beq
M_D \simeq M_{D^*} \; , \; M_B \simeq M_{B^*}.
\eeq
Simple scaling laws concerning the approach to the asymptote apply, where
\bea
M_{B^*} - M_B &\sim& \frac{m_c}{m_b} ( M_{D^*} - M_D ), \\
M_B - M_D &\sim & m_b - m_c.
\eea
It is obvious already from the spectroscopy results that beauty hadrons are heavy flavor; however,
charm hadrons also primarily act as heavy-flavor particles.

For the heavy quark expansion (HQE), one requires dimensionless parameters to define the landscape, i.e., of the order of the ratio
$\bar \Lambda/m_Q$, where $\bar \Lambda$ defines the short- vs.
long-distance dynamics in heavy-flavor decays in QCD. $\bar \Lambda$ is usually also applied in LQCD analyses with ${\cal O}(1)$ GeV (or more). This depends on the case to which it is applied. Furthermore, subtle points should be made regarding the definition of
quark masses: one uses the ``running" mass $m_Q(\mu)$, defined at a scale of $\mu$
to shield it against strong infrared dynamics. One must use ``well-defined" masses for decays, and
{\em not} pole masses. However, we require additional tools.

\subsubsection{Operator product expansion }
\label{OPE}

Operator product expansion ({\bf OPE}) ({\em \`{a} la} Wilson \cite{WILSON}
\footnote{I emphasize that there are subtle points that should be considered regardless of whether one discusses OPE in general or {\em \`{a} la} Wilson.}) provides a powerful theoretical tool of wide applicability.  
\begin{itemize}
\item
One defines a field theory ${\cal L}(\Lambda _{\rm UV})$ at a high ultraviolet scale
$\Lambda _{\rm UV}$, which is significantly higher than $M_W$, $m_Q$, etc.
\item
One renormalizes the ${\cal L}$ from the cutoff $\Lambda _{\rm UV}$ down to
the physical scale $\Lambda_{\rm phys}$ for application. In doing so, one integrates out
the {\em heavy} degrees of freedom. That is, with like $M_W$ one arrives at an
{\em effective low-energy} field theory using OPE, where
\beq
{\cal L}(\Lambda _{\rm UV}) \to {\cal L}(\Lambda _{\rm phys})=
\sum_i c_i(\Lambda_{\rm phys},\Lambda_{\rm UV},M_W,...)  {\cal O}_i (\Lambda_{\rm phys} ).
\eeq
The {\em local} operators ${\cal O}_i (\Lambda_{\rm phys})$ contain the {\em active} dynamical
fields; i.e., those with frequencies below ${\cal O}_i (\Lambda_{\rm phys})$.

\item
Their coefficients $c_i(\Lambda_{\rm phys},\Lambda_{\rm UV},M_W,...) $ provide the gateway
for heavy degrees of freedom with frequencies above ${\cal O}_i (\Lambda_{\rm phys})$ to enter.
They are shaped by short-distance dynamics and are usually computed perturbatively.

\item
Lowering the value of ${\cal O}_i (\Lambda_{\rm phys})$ changes the ``shape" of the Lagrangian, such that
${\cal O}_i (\Lambda^{(1)}_{\rm phys}) \neq {\cal O}_i (\Lambda^{(2)}_{\rm phys})$ for
$\Lambda^{(1)}_{\rm phys} \neq \Lambda^{(2)}_{\rm phys}$. Integrating out heavier fields
will induce higher-dimensional operators to emerge in the Lagrangian.

\item
As a matter of principle, observables {\em cannot} depend on the choice of
$\Lambda _{\rm phys}$. They provide a demarcation line only, with
\beq
{\rm short \; distance} \; \;  < \; \; 1/\Lambda_{\rm phys} \; \; <\; \; {\rm long \; distance}.
\eeq
In practice, the value of $\Lambda _{\rm phys}$  must be chosen judiciously, because of the present limitations of our computational powers.
It is reasonable to pick $\Lambda_{\rm phys}= \bar \Lambda \sim 1$ GeV, for application to charm transitions in particular.

\end{itemize}
We require additional \& subtle steps for inclusive weak decays. One describes the decays into sufficiently
inclusive final states in the weak interactions, using the imaginary part of the forward scattering operator up to second order accuracy and invoking the optical theorem. Thus,
\beq
T(Q \to Q) = {\rm Im}\int d^4x\;  i\{ {\cal L}(x){\cal L}(0)\}_t \; ,
\label{OPTH}
\eeq
with the subscript $t$ denoting the time-ordered product and ${\cal L}_W$ the relevant weak
Lagrangian. $T(Q \to Q)$ represent, in general, a non-local operator. The
space-time separation $x$ is given by the inverse of the {\em energy release}.
If the latter is large compared to typical hadronic scales, the product is dominated by short-distance
dynamics and one can apply an OPE. This yields an infinite series of local operators of 
increasing dimensions.

We take the $H_Q$ expectation values of the operator $T$ normalized by $2M_{H_Q}$, such that
\bea
\frac{\langle H_Q| {\rm Im}\, T(Q\to Q) | H_Q\rangle }{2 M_{H_Q}} &\propto& \\
\propto \Gamma (H_Q \to f) &=& \frac{G_F^2m_Q^5(\omega)}{192 \pi ^3} |V_{\rm CKM}|^2
\cdot \\
\cdot [c_3^{(f)}(\omega ) \frac{\langle H_Q|\bar QQ|H_Q\rangle_{(\omega)}}{2M_{H_Q}} &+&
\frac{c_5^{(f)}(\omega )}{m_Q^2} \frac{\langle H_Q|\bar Q\frac{i}{2} \sigma \cdot
G Q|H_Q\rangle_{(\omega )}}{2M_{H_Q}}    + \\
&+&\sum _i\frac{c_{6,i}^{(f)}(\omega )}{m_Q^3}
\frac{\langle H_Q|(\bar Q\Gamma_i q)(\bar q\Gamma_i Q|H_Q\rangle_{(\omega)}}{2M_{H_Q}}
+ {\cal O}(1/m_Q^4)       ].
\eea
One uses $\Lambda_{\rm phys} \ll \omega \ll m_Q$ for expansion to deal with the impact
of perturbative \& non-perturbative QCD. Short-distance dynamics shape the number of coefficients $c_i^{(f)}$. In practice, they are evaluated in perturbative QCD; they also provide the portals for ND entering naturally. Non-perturbative contributions enter through the
expectation values of operators with dimensions of five \& higher, i.e.,
$\bar Q\frac{i}{2} \sigma \cdot G Q$, $(\bar Q\Gamma _i q)(\bar q \Gamma_i Q)$, etc.
Expanding the expectation value of the leading operator $\bar QQ$ of dimension three, we obtain
\bea
\frac{1}{2M_{H_Q}} \langle H_Q|\bar QQ | H_Q\rangle _{(\omega)} &=& 1 -
\frac{\mu^2_{\pi}(\omega)}{2m_Q^2} + \frac{\mu_G(\omega)}{2m_Q^2}
+ {\cal O}(1/m_Q^3), \\
\mu^2_{\pi} (\omega) &=&
\frac{1}{2M_{H_Q}}\langle H_Q|\bar Q\vec\pi^2  Q |H_Q\rangle _{(\omega)},  \\
\mu^2_G (\omega)&=&\frac{1}{2M_{H_Q}}
\langle H_Q|\bar Q\frac{i}{2} \sigma \cdot G Q|H_Q\rangle_{(\omega )}.
\label{BARQQ}
\eea
Observables cannot depend on the value of $\omega$. A crucial difference exists between
amplitudes as, in real quantum field theories (QFT), it is not trivial to connect short- \& long-distance dynamics. However, we do have the tools to accomplish this, as it is possible to discuss
uncertainties only {\em inside} models of strong forces.

Inclusive transitions can be described in the $\bar \Lambda/m_Q$ expansion. We have learned that
{\em inclusive} transitions begin only at the {\em second order} in general, for subtle
reasons \cite{BUV},
which are related to lifetimes and semi-leptonic decays in particular. There are five points to note:
\begin{itemize}
\item
For heavy flavor hadrons, the leading source of inclusive transitions is parton models
in smart ways.
\item
Non-perturbative dynamics enter to the second order of $\bar \Lambda/m_Q$ only
(also in smart ways).
\item
For the landscape of $H_Q$ transitions, we have the same list of operators: $\bar QQ$,
$\bar Q \frac{i}{2}\sigma \cdot G Q$, $(\bar Q\Gamma _i q)(\bar q\Gamma_i Q)$, etc., for the widths and distributions. However, their
impact is very different due to subtle effects \cite{BSUV}. This is effective for the widths of beauty
and charm hadrons, but not for charm hadron distributions.
\item
HQE functions significantly better than previously expected (again in smart ways).
\item
There is a large difference between inclusive and exclusive transitions. We expect this difference,
but it is barely within our control.

\end{itemize}
The above makes sense for beauty decays, but what of $\bar \Lambda/m_c$? Obviously, it depends on the heavy quark mass. Note that one {\em cannot} use ``pole mass" because of ``old renormalon" uncertainties
\cite{BSUVPOL}.
One must use a very effective definition, called ``kinetic" mass \cite{KIN,BSUVPOL}, where
\beq
\frac{dm_Q^{\rm kin}(\omega)}{d\omega} = - \frac{16\alpha _S(\omega)}{3\pi} -
\frac{4\alpha_S}{3\pi}\frac{\omega}{m_Q} + ...,
\eeq
with a scale of $\sim 1$ GeV;
this functions very well, including at least third \& fourth order results.  We require a little luck for application to
charm hadrons; in poetic terms, we can accomplish it with ``undue incantation". Actually, the connection with lattice QCD studies gives us novel information concerning underlying fundamental dynamics, which is being tested now and will continue to be examined in the future.

The leading non-perturbative corrections arise at ${\cal O}(1/m_Q^2)$ and differentiate between baryons on one side and mesons on the other; the latter have practically the same
values. In ${\cal O}(1/m_Q^3)$, the landscapes also differentiate between mesons
with dimension six operators. One describes
Pauli interferences ({\bf PI}) that are negative for mesons, but not for baryons in general. Weak annihilation/exchanges ({\bf WA}) have a sizable impact on
baryon amplitudes. We have acquired information even from the ${\cal O}(1/m_Q^4)$ contribution and have made some estimates concerning the
${\cal O}(1/m_Q^5)$ case. One must have realistic expectations regarding charm decays.

\subsection{Sum rules and dispersion relations}
\label{OTHERS}

Other less ``famous" theoretical tools exist. They are also important and will be even more
so in the future, when we will be forced to focus on accuracy.

\subsubsection{Sum rules}

``Sum rules" are ubiquitous tools in many branches of physics, involving sums or integrals
over observables such as rates \& their moments, etc. A celebrated case is the SVZ QCD
sum rules named after Shifman, Vainshtein, \& Zakharov \cite{SVZ}, which allow
low-energy hadronic quantities to be expressed through basic QCD parameters. An {\bf OPE} is obtained, 
and non-perturbative dynamics are then parametrized through
condensates $\langle 0|\bar q q |)\rangle$, $\langle 0 |GG|0\rangle$, etc. They are zero in
perturbative QCD; however, they are treated as free parameters, the values of which are
fitted from certain observables. This approach also indicates that the duality between the worlds of the hadrons \& quarks (\& gluons) is not always local, i.e., we must treat ``smeared" hadronic observables.  
The first real example is the description of $e^+ e^- \to \gamma ^* \to$ hadrons in the
energy range $E_{c.m.} \sim 3.6-5$ GeV, including narrow resonances.

One can also apply ``light-cone sum rules" \cite{LSC}, ``small velocity (SV)" \cite{SV}, and
``spin sum rules" \cite{SSR}.  Certain examples exist to which the {\bf OPE} is applicable:
\bea
\mu ^2_G(\omega ) &\leq& \mu _{\pi}^2(\omega),   \\
\mu ^2_G(1\, {\rm GeV}) &\simeq& \frac{3}{2} [M^2_{B^*} - M^2_B ] \simeq 0.35 \pm 0.03 \;
({\rm GeV})^2, \\
\mu _{\pi}^2(1\; {\rm GeV}) &\simeq &0.45 \pm 0.1 \; {\rm (GeV)^2}.
\eea
The {\bf OPE} is applicable through ${\cal O}(1/m_Q^3, 1/m_Q^4)$ (and more \cite{SASHA}) for beauty decays.
This approach is also surprisingly applicable to charm transitions, through ${\cal O}(1/m_c^3)$.

\subsubsection{Dispersion relations}

Dispersion relations \cite{KUBIS,DR} are encountered in many branches of physics and in quite different
contexts; they are based on the general validity of central statements in QFT. We can relate
the values of a two-point function $\Pi (q^2)$ in a QFT at different complex values of
$q^2$ to each other through an integral representation. In particular, one can evaluate
$\Pi (q^2)$ for large Euclidean values with the help of an {\bf OPE}, and then relate the
coefficients $I_n^{\rm OPE}$ of local operators $O_n$ to observables such as
$\sigma (e^+e^- \to {\rm hadrons})$ \& their moments in the physical, i.e., Minkowskian domain. This is achieved
by taking an integral over the discontinuity around the real axis, such that
\beq
I_n^{\rm OPE} \simeq \frac{1}{\pi} \int_0^{\infty} ds \frac{s}{(s+q^2)^{n+1}} \cdot \sigma (s) \; .
\eeq
The integral over the asymptotic arcs vanishes.

Those results are based on physical singularities, poles, \& cuts only on the
real axis of $q^2$. This is the basis of the derivation of the celebrated QCD sum rules \cite{SVZ}. Such dispersion relations are used to calculate transition rates
in the {\bf HQE}, and to derive new classes of sum rules such as those given in \cite{SV}.

\subsection{Probing three-body FS}

The usual Breit-Wigner parametrization does not describe the impact of broad resonances
such as $\sigma/f_0(500)$ \& $\kappa/K^*_0(800)$ \cite{KUBIS,DR} on both charm (\& beauty) hadronic FS well, for various reasons. The interference of narrow and broad resonances cannot be described as being simply ``inside" and ``outside" the centers of the narrow resonances. Instead, they must be described in a more subtle manner,
i.e., in relation to fractional asymmetries, significance, etc. \cite{REIS,WILL}.
Again, this depends on the situation. However, comparing results provides us with information
about non-perturbative QCD at least.
We have the tools to probe the two-dimensional Dalitz plots, with a long history in
strong dynamics in particular.
One requires larger amounts of data and more extensive experimental work,
but ``rewards" are also available, specifically, information about the existence of ND and its
features. One can use model insensitive analyses as the second step.  Ultimately, these technologies must agree following thought \& discussion; at minimum, they will provide us with information concerning strong forces.

We require a third step at minimum. {\bf FSI} by strong forces cannot be calculated from first principles at present. Yet, one can relate these factors using non-trivial
theoretical tools incorporating chiral symmetry and refined dispersion relations \cite{KUBIS,DR}, which are
based on data concerning low-energy pion and kaon collisions. The crucial strength in this approach is that we cannot
depend on the best fitted data, but rather on correlations with other transitions based on tested theories.

\subsection{Four-body FS with different roads to ND}

When we measure four-body FS, we must engage with the three-dimensional world, i.e.,
complex situations. Then, one must be both realistic \& clever.
In this scenario, {\bf FSI} have even more impact as regards
changing the worlds of quarks vs. hadrons. The landscapes of four hadrons in the FS are very different for several reasons, some are which are obvious, while others are more subtle. Therefore, one must both consider and attempt different approaches to probing CP asymmetries in four-body FS. Furthermore, our goal is to show the impact of SM vs. ND.

A comparison between {\em T-odd moments} of $H_Q \to h_1h_2h_3h_4$ vs.
$\bar H_Q \to \bar h_1\bar h_2\bar h_3 \bar h_4$ in a center-of-mass frame was suggested, with
$\langle {\rm A}_{\rm T \, odd}\rangle = \langle \vec p_1 \cdot (\vec p_2 \times \vec p_3)  \rangle$
for $H_Q$ decays vs. $\langle {\rm \bar A}_{\rm T \, odd}\rangle = \langle \vec {\bar p_1} \cdot
(\vec {\bar p_2} \times \vec {\bar p_3})  \rangle$ for $\bar H_Q$ decays, leading to CP asymmetry
 \cite{PISA}. Then,
$\langle A_{\rm CPV}\rangle =
\frac{1}{2}[{\langle \rm A}_{\rm T \, odd}\rangle  - \langle \bar {\rm A}_{\rm T \, odd}\rangle ]$.
Later, this approach was discussed in more detail for beauty mesons \& baryons \cite{DLONDON}, and was in fact suggested for special situations such as $B \to VV$  \cite{VALENCIA} at an earlier stage.
This is an intelligent approach to measuring asymmetries independent of production asymmetries, and has been used with real data \cite{FOCUS,BABARODD,LHCbODD} in the case of charm mesons. The
definitions $C_T \equiv \vec p_1 \cdot (\vec p_2 \times \vec p_3)$ and
$\bar C_T \equiv \vec {\bar p_1} \cdot (\vec {\bar p_2} \times \vec {\bar p_3})$ lead to
{\em T-odd} observables, where
\beq
A_{T} \equiv \frac{\Gamma_{H_Q}(C_T >0) - \Gamma_{H_Q}(C_T <0)}
{\Gamma_{H_Q}(C_T >0) + \Gamma_{H_Q}(C_T <0)} \; , \;
\bar A_{T} \equiv \frac{\Gamma_{\bar H_Q}( \bar C_T < 0) -
\Gamma_{\bar H_Q}( \bar C_T >0)}
{\Gamma_{\bar H_Q}(\bar C_T <0) + \Gamma_{\bar H_Q}( \bar C_T >0)}.
\eeq
FSI can produce $A_T$, $\bar A_T$ $\neq 0$ without CPV; yet, with non-zero difference, one establishes
CP asymmetry
\beq
a_{CPV}^{T-odd} \equiv \frac{1}{2} (A_T - \bar A_T ) \; .
\eeq
With more data \& additional thought, we may develop some ideas concerning a ``better"
value for $d >0$ that does not depend on experimental findings only.
\beq
A_{T} (d) \equiv \frac{\Gamma_{H_Q}(C_T >d) - \Gamma_{H_Q}(C_T <-d)}
{\Gamma_{H_Q}(C_T >d) + \Gamma_{H_Q}(C_T <-d)} \; , \;
\bar A_{T} (d)\equiv \frac{\Gamma_{\bar H_Q}(\bar C_T < -d) -
\Gamma_{\bar H_Q}(\bar C_T > d)}
{\Gamma_{\bar H_Q}(\bar C_T < -d) + \Gamma_{\bar H_Q}(\bar C_T >d)}.
\eeq

However, we {\em cannot} stop there. We require one-dimensional observables, although we also require additional data, subtle analyses, and thought. For example, one can measure the angle
between two planes \cite{CPBOOK, TAUD+,YOGI}, where 
\bea
\frac{d\Gamma}{d\phi} (H_Q \to h_1h_2h_3h_4) &=& \Gamma_1 {\rm cos^2}\phi + \Gamma_2 {\rm sin^2}\phi +\Gamma_3 {\rm cos}\phi {\rm sin}\phi,
\\
\frac{d\Gamma}{d\phi} (\bar H_Q \to \bar h_1 \bar h_2 \bar h_3 \bar h_4) &=& \bar \Gamma_1 {\rm cos^2}\phi + \bar \Gamma_2 {\rm sin^2}\phi -\bar \Gamma_3 {\rm cos}\phi {\rm sin}\phi.
\eea
Integrated rates give $\Gamma_1+\Gamma_2$ vs. $\bar \Gamma_1 + \bar \Gamma_2$, where
\beq
\Gamma (H_Q \to h_1h_2h_3h_4) = \frac{\pi}{2} (\Gamma_1 + \Gamma_2)  \; \; \; {\rm vs.} \; \; \;
\Gamma (\bar H_Q \to \bar h_1\bar h_2 \bar h_3 \bar h_4)
= \frac{\pi}{2} (\bar \Gamma_1 + \bar \Gamma_2)  \; .
\eeq
$\Gamma _3$ \& $\bar \Gamma _3$ can be compared with
published $\langle A_{\rm T\, odd}\rangle$ \& $\langle \bar A_{\rm T\, odd}\rangle$, as discussed above \cite{CPBOOK}; this shows the already sizable impact of re-scattering. The moments of
{\em integrated forward-backward} asymmetry
\beq
\langle A_{\rm FB}\rangle =
\frac{\Gamma_3 - \bar \Gamma_3}{\pi(\Gamma_1+\Gamma_2+\bar \Gamma_1+\bar \Gamma_2)},
\eeq
provide information about CPV. This can be tested \& compared as used by
\cite{FOCUS,BABARODD,LHCbODD}, and as shown in Sect.\ref{CPVMES} below.

In the future, we should probe semi-regional asymmetries. One could also disentangle $\Gamma_1$ vs. $\bar \Gamma_1$ and $\Gamma_2$ vs. $\bar \Gamma_2$ by tracking the distribution in the angle $\phi$; $\Gamma_1 \neq \bar \Gamma_1$
and/or  $\Gamma_2 \neq \bar \Gamma_2$ represent direct CPV in the partial width.

If there is a {\em production} asymmetry, it gives global
$\Gamma_1 = c \bar \Gamma_1$, $\Gamma _s = c \bar \Gamma _2$,
and $\Gamma_3 = - c \bar \Gamma_3$ with {\em global} $c \neq 1$. Furthermore, one can apply these observables to different definitions of those planes (as discussed below) and their correlations. This will help us to understand these underlying forces.

There are subtle methods of defining $\phi$.
We have learned from the history surrounding
$K_L \to \pi^+\pi^- \gamma ^*  \to \pi^+\pi^- e^+e^-$,
where Seghal \cite{SEGHAL1} predicted CPV of approximately 14 \% based on
$\epsilon_K \simeq 0.002$. Unit vectors aid in discussing this scenario in more detail, where
\beq
\vec n _{\pi} = \frac{\vec p_+ \times \vec p_-}{|\vec p_+ \times \vec p_-|} \; , \;
\vec n _{l} = \frac{\vec k_+ \times \vec k_-}{|\vec k_+ \times \vec k_-|}  \; , \;
\vec z = \frac{\vec p_+ + \vec p_-}{|\vec p_+ + \vec p_-|}, \\
\eeq
\bea
{\rm sin} \phi = ( \vec n _{\pi}  \times \vec n _{l} ) \cdot \vec z \; [CP=-,T=-] &,&
{\rm cos} \phi = \vec n _{\pi}  \cdot \vec n _{l} \; [CP=+,T=+],\\
\frac{d\Gamma}{d\phi} &\sim & 1 -(Z _3\,  {\rm cos} 2\phi  + Z_1\, {\rm sin}2\phi).
\eea
Then, one measures asymmetry in the moments
\beq
{\cal A}_{\phi} =
\frac{(\int_0^{\pi/2} - \int_{\pi/2}^{\pi}+\int_{\pi}^{3\pi/2}-\int_{3\pi/2}^{2\pi})\frac{d\Gamma}{\phi}}
{(\int_0^{\pi/2} + \int_{\pi/2}^{\pi}+\int_{\pi}^{3\pi/2}+\int_{3\pi/2}^{2\pi}) \frac{d\Gamma}{\phi}}.
\eeq
There is an obvious reason for probing the angle between the two
$\pi^+\pi^-$ \& $e^+e^-$ planes only, which is based on $K_L \to \pi^+\pi^- \gamma ^*$
or $K^0 \to \pi^+\pi^- \gamma ^*$ vs. $\bar K^0 \to \pi^+ \pi^- \gamma^*$.

However, these situations are more complex, as
\bea
\frac{d}{d\phi} \Gamma (H_Q \to h_1h_2h_3h_4) &=&
|c_Q|^2 - [b_Q\,  (2 {\rm cos}^2 \phi  -1) + 2a_Q\, {\rm sin}\phi \, {\rm cos}\phi],  \\
\frac{d}{d\phi} \Gamma (\bar H_Q \to \bar h_1\bar h_2\bar h_3\bar h_4) &=&
|\bar c_Q|^2 - [\bar b_Q\,  (2 {\rm cos}^2 \phi  -1) - 2\bar a_Q\, {\rm sin}\phi \,
{\rm cos}\phi], \\
\Gamma (H_Q \to h_1h_2h_3h_4) = |c_Q|^2 \; \; \;  &{\rm vs.}& \; \; \;
\Gamma (\bar H_Q \to \bar h_1\bar h_2 \bar h_3\bar h_4) = |\bar c_Q|^2,  \\
\langle A_{\rm CPV}^Q \rangle &=& \frac{2(a_Q - \bar a_Q)}{|c_Q|^2+|\bar c_Q|^2}  \; ,
\eea
i.e., the $b_Q$ \& $\bar b_Q$ terms have no impact.
Furthermore, one wishes to probe semi-regional asymmetries,
to which $b_Q$ and $\bar b_Q$ contribute, with
\beq
A_{\rm CPV}^Q|_e^f =
\frac{\int _e^f d\phi \frac{d\Gamma}{d\phi}- \int_e^f d\phi\frac{d\bar \Gamma}{d\phi}}
{\int _e^f d\phi\frac{d\Gamma}{d\phi}+ \int_e^f d\phi\frac{d\bar \Gamma}{d\phi}} \; .
\eeq
Again, one should not choose which approach gives the best fitting result, but should instead follow a deeper reasoning.

These examples are correct as regards the general theoretical bases. However, some
are more successful as regards experimental uncertainties, cuts, and/or probing the impact of ND. Also, the true underlying dynamics do not produce the best fitting of the data.
Furthermore, it is crucial to use CPT invariance as a tool for correlations with other transitions.

\subsection{A very short summary}

It is important to learn about theoretical tools, and particularly their correlations with each other.
{\bf OPE}, {\bf HQE}, sum rules, dispersion relations, and LQCD are important now, and will be in the future when more consideration
\& deliberation is given to the connection between charm \& beauty hadrons. These connections depend on where
and how. Charm transitions show us the meaning of
``charming a cobra," as regards the manner in which theorists can use them in their calculations. At least 
charm quarks are mostly on the ``right side".

\section{Leptonic, semi-leptonic, \& rare charm decays}
\label{LEPTON}

A rich landscape for learning about fundamental dynamics is provided by (semi)leptonic decays of charm hadrons, but I will focus on two items. These relate to a possible sign
of ND and the development of an improved understanding of strong spectroscopy.

\subsection{Leptonic decays of $D^+$ and $D_s^+$}
The landscapes of leptonic decays of $D^+_q \to l^+ \nu (+\gamma 's) $ with $q=d,s$ and
$l= \tau, \mu, e$ are less complex. The SM predictions depend on two parameters in the amplitudes, namely, $|V_{cq}|$ (due to weak forces) and $f_{D_q}$ (due to non-perturbative QCD).
The amplitudes are given with $W^+$ exchanges by
\bea
{\cal T} (D^+_q \to l^+\nu) &=& \frac{G_F}{\sqrt 2} \langle 0|A^{\mu} |D_q \rangle
[\bar l \gamma _{\mu} (1-\gamma_5)\nu_l ] \\
\langle 0|A^{\mu}|D_q(p)\rangle    &=& if_{D_q}p^{\mu} \; , \;
A^{\mu} = \bar c \gamma ^{\mu}(1-\gamma_5)q,  \\
\Gamma (D^+_q \to l^+\nu_l)      &=& \frac{G_F^2}{8\pi} |f_{D_q}|^2|V_{cq}|^2m^2_l
\left(1-\frac{m_l^2}{M_{D_q}^2} \right)^2 M_{D_q} 
\eea
The SM prediction shows the impact of chiral symmetry on the amplitude with $m_l$
\footnote{I also list
BR$(D^+ \to e^+ \nu_l) = 1.07 \cdot 10^{-8} \cdot (f_D/220\, {\rm MeV})^2$ and
BR$ (D^+_s \to e^+ \nu_l) = 1.2 \cdot 10^{-7} \cdot (f_{D_s}/250\, {\rm MeV})^2$,
and compare the experimental limits. Hence, BR$(D^+ \to e^+ \nu_l) \leq  8.8 \cdot 10^{-6}$ and
BR$(D^+_s \to e^+ \nu_l) \leq  8.3 \cdot 10^{-5}$. Is this a hopeless enterprise?
``Miracles" can happen.}: 
\bea
{\rm BR} (D^+ \to \tau^+ \nu_l) &=& 1.0 \cdot 10^{-3} \cdot (f_D/220\, {\rm MeV})^2,   \\
{\rm BR} (D^+ \to \mu^+ \nu_l) &=& 4.6 \cdot 10^{-4} \cdot (f_D/220\, {\rm MeV})^2,\\
{\rm BR} (D^+_s \to \tau^+ \nu_l) &=& 4.5 \cdot 10^{-2} \cdot (f_{D_s}/250\, {\rm MeV})^2,\\
{\rm BR} (D^+_s \to \mu^+ \nu_l) &=& 5.0 \cdot 10^{-3} \cdot (f_{D_s}/250 \, {\rm MeV})^2.
\eea
It is a well-known fact that $f_D$ \& $f_{D_s}$  provide us with very good tests of our quantitative control over non-perturbative QCD, through LQCD and, to an even greater extent, the $f_D/f_{D_s}$ ratio.

The data are consistent with these predictions, but they leave sizable space for ND, in
particular in relation to charged Higgs exchanges, where
\bea
{\rm BR} (D^+ \to \tau^+ \nu) \leq  1.2 \cdot 10^{-3} &,&
{\rm BR} (D^+ \to \mu^+ \nu) = (3.82\pm 0.33) \cdot 10^{-4} \\
{\rm BR} (D^+_s \to \tau^+ \nu) = (5.54\pm 0.24) \cdot 10^{-2} &,&
{\rm BR} (D^+_s \to \mu^+ \nu) = (5.56 \pm 0.25) \cdot 10^{-3} 
\eea
Could this possibly constitute an indirect gateway for ``Dark Matter"?

\subsection{Exclusive semi-leptonic decays of charm mesons}

There are several excellent reasons for measuring exclusive semi-leptonic $D^+_{(s)}$ decays
with accuracy. Here, I will comment on only one item, i.e., the spectroscopy of neutral
mesons, its connections with weak dynamics, and testing these data with radiative
decays as discussed in detail in Ref.\cite{DON} (\& the long list of references). In quark models,
we describe wave functions of the
$I=0$ neutral $\eta$ and $\eta^{\prime}$ as $|\bar q_iq_i\rangle$ states including mixing
\footnote{The term ``mixing" covers broader items than ``oscillations"; the latter can be applied
to neutral meson (or $N-\bar N$) transitions only and, crucially, it depends on the impact of ``time".}
between $|\eta _8\rangle = \frac{1}{\sqrt 6}|\bar u u +\bar d d -2\bar ss\rangle$
and $|\eta _0\rangle = \frac{1}{\sqrt 3}|\bar u u +\bar dd +\bar ss\rangle$. With
non-perturbative QCD we must discuss the impact of {\em another} $I=0$ neutral state, such as
$|gg\rangle$ with ``constituent" gluons.

The $|\eta\rangle$ \& $|\eta^{\prime}\rangle$ states act as initial \& final states and also in between. 
This increases the complexity of light meson spectroscopy
\footnote{One might put $Z_{\eta } \simeq 0$ assuming that $|\eta ^{\prime}\rangle$ contains more
gluonic components.}, with
\bea
|\eta^{\prime}\rangle &=& X_{\eta^{\prime}} |\eta_0\rangle +
Y_{\eta^{\prime}} |\eta_8\rangle + Z_{\eta^{\prime}} |gg \rangle, \\
|\eta \rangle &=& X_{\eta} |\eta_0\rangle +
Y_{\eta} |\eta_8\rangle + Z_{\eta } |gg \rangle.
\eea
Gluonic components change the information
we can obtain from $D^+_{(s)} \to l^+ \nu \eta$, $l^+\nu\eta^{\prime}$ (with $l=e, \mu$) providing data
concerning non-perturbative QCD. One can continue with
$B^+ \to l^+ \nu \eta$, $l^+ \nu \eta^{\prime}$, where $l$ includes $\tau$.
Furthermore, we have a non-zero chance of finding the sign of ND, in particular as regards
$B^+ \to \tau^+ \nu \eta$, $\tau^+ \nu \eta^{\prime}$. Even more ambitious researchers can
use these tools to probe exclusive cases such as $D^+\to \pi^+\eta/\eta^{\prime}$,
$B^+ \to \pi \eta/\eta^{\prime}$, etc. This is not only a theoretical consideration of the connections
between strong spectra and exclusive weak decays. We have tested
these connections with electromagnetic dynamics, and accurately treated:
$\psi^{\prime}, \psi, \phi \to \gamma \eta ^{\prime}$ vs. $\gamma \eta$;
$\rho, \omega \to \gamma \eta$; $\eta ^{\prime} \to \gamma \omega , \gamma \rho$;
$\eta^{\prime} \to 2 \gamma$ vs. $\eta \to 2 \gamma$;
$\gamma \gamma \to \eta$ vs. $\gamma \gamma \to \eta^{\prime}$;
$\psi \to \rho/\omega/\phi + \eta$ vs. $\psi \to \rho/\omega/\phi + \eta^{\prime}$, etc.
However, following these discussions and analyses of the obtained data, we have not yet reached the final conclusions.

\subsection{Rare decays}
\label{RARE}

Rare decays of beauty (\& strange) hadrons provide us with a deeper understanding of fundamental
dynamics. However, the scenarios are very different for charm transitions: long-distance strong forces 
are very important (or more), over which we have little control. First, one can discuss very
rare decays: 
\bea
{\rm BR}(D^0 \to 2\gamma) |_{\rm exp} &\leq & 2.2 \cdot 10^{-6},\\
{\rm BR}(D^0 \to \mu^+\mu^-) |_{\rm exp} &\leq & 6.2 \cdot 10^{-9}.
\eea
Guesstimates using the second-order GIM effect, helicity suppression, \&
$\frac{f_D}{m_c} \ll 1$ yield
\beq
{\rm BR}(D^0 \to \mu^+\mu^-) \sim {\cal O} \left( {\rm BR}(D^+ \to \mu^+\nu) \cdot \frac{\alpha_S}{\pi}
\cdot \frac{m_s^2}{M_W^2} \right) \sim {\cal O}(10^{-12}).
\eeq
A more detailed treatment is provided by the SM \cite{SINGER,BURD}, with
\bea
{\rm BR}(D^0 \to 2\gamma) &\sim& (1-3.5) \cdot 10^{-8}, \\
{\rm BR}(D^0 \to \mu^+\mu^-) &\sim& 2.7 \cdot 10^{-5} \cdot {\rm BR}(D^0 \to 2\gamma)
\sim (0.3 - 1) \cdot 10^{-12}.
\eea
Theoretical tools for refined analyses of the SM based on OPE and including long-distance
dynamics with quark condensates exist. However, this would be considered as an academic exercise in view of the very
tiny rates.

On the positive side, one can search for manifestations of ND \cite{BURD,PPB} in a wide range, where
\beq
{\rm BR} (D^0 \to \mu^+\mu^-) |_{\rm ND} \sim 10^{-11}/10^{-10}/8\cdot 10^{-8}/3.5 \cdot 10^{-6},
\eeq
with superheavy $b^{\prime}$ quark/``warped extra dimension"/multi-Higgs sector/supersymmetry (SUSY) with $R$
parity breaking \footnote{Of course, SUSY with $R$ parity breaking can do almost anything.}.

I have referred very indirectly to the ``strong CP challenge" regarding the ABJ anomaly. The effective Lagrangian for the strong forces is expressed as
${\cal L}_{\rm eff} = {\cal L}_{\rm QCD} + \frac{\theta g_S^2}{32 \pi ^2}G\cdot \tilde G$.
Limits given by the data indicate that
``un-natural" $\theta < 10^{-9}$ \footnote{Including weak decays, this shows that observable
dynamics depend on the combination $\bar \theta = \theta - {\rm arg\, det}{\cal M}$ with the
quark mass matrix ${\cal M}$.}. To make this ``natural", it has been suggested that
Peccei-Quinn symmetry should be introduced \cite{PQS}. This implies the existence of ``axions", which have been elusive to date. Regardless, ``familons" can be their flavor-nondiagonal partners. We have not found
axions in $K^+ \to \pi^+ f^0$, $B^+ \to \pi^+/K^+ f^0$, or $B_d \to K_Sf^0$; however, no real limits have been established in $D^+ \to \pi^+/K^+ f^0$ as of yet.

Less rare decays include inclusive $D_{(s)} \to \gamma X_q$, $D_{(s)} \to l^+l^- X_q$, \&
exclusive $D_{(s)} \to \gamma K^*/\rho/\omega/\phi$, $D_{(s)} \to l^+l^- K^*/\rho/\omega/\phi$
or $\Lambda_c^+ \to \gamma p$,$l^+l^-P$, etc. The main problem is
that long-distance strong forces can produce rates like present limits, with
$D \to \gamma K^*/\rho/\omega$, $l^+l^-K^*/\rho/\omega$ like $D^0 \to \rho \rho \to l^+l^-\rho$,
etc. SM provides order-of-magnitude predictions, with typical numbers being \cite{BURD2}
\bea
{\rm BR}(D^0 \to \gamma \bar K^{*0}) = (6-36) \cdot 10^{-5}\; \;  &,&\; \;
{\rm BR}(D^0 \to \gamma \rho ^0) = (0.1-1) \cdot 10^{-5}, \\
{\rm BR}(D^0 \to \gamma \omega ) = (0.1-0.9) \cdot 10^{-5}\; \;  &,&\; \;
{\rm BR}(D^0 \to \gamma \phi) = (0.1-3.4) \cdot 10^{-5}.
\eea
Present data yield
\bea
{\rm BR}(D^0 \to \gamma \bar K^{*0}) = (3.27 \pm 0.34) \cdot 10^{-4}  &,&
{\rm BR}(D^0 \to \gamma \rho ^0)\leq 2.4 \cdot 10^{-4}, \\
{\rm BR}(D^0 \to \gamma \omega) \leq 2.4 \cdot 10^{-4} &,&
{\rm BR}(D^0 \to \gamma \phi) = (2.70 \pm 0.35) \cdot 10^{-5}.
\eea
These numbers indicate that there is little reason for celebration regarding achievements on both the theoretical and experimental sides of this research field.
Future data may provide us with lessons about non-perturbative QCD. To be realistic, rates cannot indicate the existence of ND. We must measure {\em regional} asymmetries such as
forward-backward and/or CP asymmetries. This is the only opportunity to probe the impact of ND,
since long-distance dynamics cannot produce these results \cite{PPB}. Therefore, we require extremely large data sets
of $D_q \to l^+l^- K^*/\rho/\omega/\phi$ or $\Lambda^+_c \to l^+l^- p$, etc.

\section{Non-leptonic decays \& CP asymmetries}
\label{NL}

There are several classes of charm (\& beauty) hadron transitions. I will focus on
inclusive decays (lifetimes \& semi-leptonic branching ratios) and CP asymmetries.
\begin{itemize}
\item
Inclusive decays test our control over non-perturbative QCD; there is
no other candidate for strong forces in local QFT.
\item
CP asymmetries are related to weak dynamics, in particular regarding the connection of $SU(2)_L \times U(1)$ with $SU(3)_C$. This topic is complex for several reasons; for example, one must
understand the dynamics between different exclusive FS and, therefore, the
quantitative impact of re-scattering.

On the positive side, the DCS landscapes are less complex because, in the world of quarks, there is only one
operator $c \to u \bar s d$. Furthermore, the SM produces almost no ``background" for CP asymmetries, which aids in our search for theimpact of ND. Of course, significantly more data is required to probe this case. 

On the other hand, this case is more complex as regards SCS: there are two (refined) tree operators
plus penguin diagrams and the SM gives small, but not zero, asymmetries. Furthermore,
the data in relation to two-body FS are closer to expectation.

Again, the ability to draw diagrams does not mean we understand the dynamics.

\end{itemize}

One uses bound states of (anti-)quarks for $D^0 = [c\bar u]$, $D^+ = [c\bar d]$,
and $D^+_s = [c\bar s]$ or, for charm baryons, $\Lambda^+_c = [c(ud)_{I=0})]$,
$\Xi_c^+ = [csu]$, $\Xi_c^0 = [csd]$, and $\Omega_c^0 = [css]$, which decay weakly. $\Sigma^{0,+,++}_c = [cdd],[c(ud)_{I=1})]$,and $[cuu]$, which decays strongly. One can compare the masses
of $\Lambda_c^+$ vs. $\Xi_c^+$ and $\Omega_c^0$ vs. $\Xi_c^0$ under U-spin symmetry
\cite{LIPKINBOOK}. Hence, one can see its violation in the differences between the ``constitute" quarks:
$m_s^{\rm const} - m_d^{\rm const}\sim 0.2$ GeV for real strong forces.

\subsection{Lifetimes and semi-leptonic decays of charm hadrons}

Equations (\ref{OPTH})--(\ref{BARQQ}) based on {\bf OPE} \& {\bf HQE} apply to Lagrangians
in general; likewise for semi-leptonic decays, where
one has ${\cal L} \propto  l_{\mu \nu} W^{\mu \nu}$ with $l_{\mu \nu}$ describing
leptonic forces and $W^{\mu \nu}$ the hadronic component. We have the tools to discuss total 
\& semi-leptonic widths for charm hadrons, but not for a discussion of the energy distributions. A comparison between data and our expectations does not yield surprising results, but this does not mean that we can truly predict those numbers quantitatively.

\subsubsection{Inclusive meson decays}
\label{INCLMES}

Careful {\bf HQE} analysis reveals that the {\bf WA} contributions are helicity suppressed and/or 
because they are non-factorizable.
On the other hand, {\bf PI} through $1/m_c^3$ occur in Cabibbo-favored $D^+$ decays;
with $f_D \propto 1/\sqrt{m_c}$ we obtain $f_D^2/m_c^2 \propto 1/m_c^3$ and, thus, semi-quantitatively \cite{CICERONE}
\bea
\Gamma (D^0) \simeq \Gamma (D_s^+) &\simeq&      \Gamma _{\rm spect} (D),  \\
\Gamma (D^+) = \Gamma _{\rm spect}(D) + \Delta \Gamma _{PI} (D^+) \; &,& \;
\frac{\tau(D^+)}{\tau(D^0)} \sim  1+ (f_D/200 \, {\rm MeV})^2 \sim 2.4 
\eea
to be compared with the data \cite{PDG14}: 
\beq
\frac{\tau (D^+)}{\tau (D^0)} = 2.54 \pm 0.01.
\eeq
The closeness of the value provided by the simple {\bf HQE} to the data is amazing. This can also be expressed in terms of
\bea
{\rm BR} (D^+ \to e^+\nu X) = (16.07 \pm 0.30) \%  &,&
{\rm BR} (D^0 \to e^+\nu X) = (6.49 \pm 0.11) \%,  \\
\frac{{\rm BR}_{\rm SL}(D^+)}{{\rm BR}_{\rm SL}(D^0)} = 2.50 \pm 0.27  \; \;  & vs. &
\; \;
\frac{\tau (D^+)}{\tau (D^0)} = 2.54 \pm 0.01.
\eea
Still, this is not the end of the road, as
\bea
\frac{\tau (D_s^+)}{\tau (D^0)} \sim  1.0 - 1.07  \; \; \; {\rm without \; WA } \; &,& \;
\frac{\tau (D_s^+)}{\tau (D^0)} \sim  0.9 - 1.3  \; \; \; {\rm with\; WA }, \\
\frac{\tau (D_s^+)}{\tau (D^0)} &=& 1.22 \pm 0.02,  \\
{\rm BR} (D^+_s \to e^+\nu X) = (6.5 \pm 0.4) \%  &,&
{\rm BR} (D^0 \to e^+\nu X) = (6.49 \pm 0.11) \%, \\
\frac{{\rm BR}_{\rm SL}(D^+_s)}{{\rm BR}_{\rm SL}(D^0)} &\sim& 1.0 \pm 0.07 \;.
\eea
Again, there is no true surprise here. However, this only means that the landscape of
non-perturbative QCD is ``subtle". For example, the ``constituent" gluons (as discussed above)
may have a role in $\eta ^{\prime}$ and $\eta$ wave functions.

It is easier to discuss ratios than absolute values. However, such numbers can provide us with additional understanding
of the underlying dynamics. A good example is provided by the lifetimes of the charm mesons \cite{PDG14}, where
\bea
\tau (D^+) &=& (1040 \pm 7) \cdot 10^{-15} \; {\rm s},  \\
\tau (D^0) = (410.1 \pm 1.5) \cdot 10^{-15} \; {\rm s}  \; &,& \;
\tau (D^+_s) = (500 \pm 7) \cdot 10^{-15} \; {\rm s}.
\eea
In the case of parton models, it has been argued that the $\tau (D^+)$ indicates the real parton tree
prediction (and also for BR$(D^+ \to l \nu X)$), while
$\tau (D^0)$ \& $\tau (D^+_s)$ can carry the impact of {\bf WA} diagrams.
However, a refined {\bf HQE} and
$m_c^{\rm kin}$ provide a different landscape, in which {\bf PI} is the leading source of the differences
and {\bf WA} is a non-leading source. After some additional subtle analyses, we can understand why the impact of
{\bf PI} is negative in meson transitions.

More refined and recent analysis is given in \cite{LENZMEMO}, yielding
\bea
\frac{\tau (D^+)}{\tau (D^0)} ^{\rm HQE2013} &=&
2.2 \pm 0.4|_{\rm hadronic} |^{+0.03}_{-0.07}|_{\rm scale}, \\
\frac{\tau (D^+_s)}{\tau (D^0)} ^{\rm HQE2013} &=& 1.19 \pm 0.12|_{\rm hadronic}
\pm 0.04|_{\rm scale}.
\eea
It is surprising that the application of {\bf HQE} to charm meson decays is so effective.
This approach will allow LQCD to be tested with other correlations in the future.

\subsubsection{Inclusive decays of charm baryons \& correlations with mesons}
\label{INCLBAR}

HQE gives predictions for charm baryon decays (all with spin $\frac{1}{2}$). At present, one requires
quark model matrix elements leading to: 
\beq
\frac{\tau (\Lambda^+_c)}{\tau (\Xi_c^0)} \sim 1.9 \;   , \;
\frac{\tau (\Xi^+_c)}{\tau (\Xi^0_c)} \sim 2.8 \; , \;
\frac{\tau (\Xi^+_c)}{\tau (\Omega_c^0)} \sim 4  \; , \;
\frac{\tau (\Xi^0_c)}{\tau (\Omega_c^0)} \sim 1.4  \; .
\eeq
Comparisons with present data yield acceptable results \cite{PDG14} 
\bea
\tau (\Lambda_c^+ =[cud]) &=& (200 \pm 6)    \cdot 10^{-15} \; {\rm s}, \\
\tau(\Xi_c^+ =[csu]) = (442 \pm 26)) \cdot 10^{-15} \; {\rm s}   &, &
 \tau(\Xi_c^0 =[csd]) = (112^{+13}_{-10}) \cdot 10^{-15} \; {\rm s}, \\
\tau (\Omega_c^0 = [css]) &=& (69\pm 12 ) \cdot 10^{-15} \; {\rm s}.
\eea
We understand why the impact of {\bf WA} on baryon decays is large, and {\bf PI} can be
either negative or positive \cite{CICERONE}.

One can predict the connection between the worlds of the mesons and baryons. One might be of the view that the scale is
given by $\frac {\tau (D^0)}{\tau (\Lambda^+_c)} \sim 2$; however, one can examine the ratio of
the longest and shortest lifetimes of charm hadrons
\beq
 \frac{\tau (D^+)}{\tau (\Omega _c^0)}\sim 13 \; ,
\eeq
noting that the data gives a factor of $\sim$ 14. It is amazing that these values are so close
considering the uncertainties in the theoretical predictions. Is this result simply ``luck"?

We have data concerning semi-leptonic decays for $\Lambda_c$ \cite{PDG14} only, where
\beq
{\rm BR}(\Lambda_c \to e^+\nu X) = (4.5 \pm 1.7) \%.
\eeq
In my view, this suggests that future data will yield a smaller value, as explained
in detail in Ref.\cite{CICERONE}. For example, one can refer to p. 82 \& Fig. 22 of this reference. These reasons for expecting a smaller value can be summarized as follows:

\noindent 
(a) {\bf PI} have a large negative impact on charm mesons, although {\bf PI} has large positive
or negative signs for baryons. Furthermore, {\bf WA} are not suppressed for baryons. Therefore, we are not
surprised (from a semi-quantitative perspective) by the data  $\frac {\Gamma (\Lambda^+_c)}{\Gamma (D^0)} \sim 2$.
Therefore, I ``predict"
BR$_{\rm SL}(\Lambda_c^+ \to e^+ \nu X) \sim \frac{1}{2} {\rm BR}_{\rm SL}(D^0 \to e^+\nu X) \sim
(3.29 \pm 0.06)\%$, which is somewhat smaller than $(4.5 \pm 1.7) \%$. 
It is based on the understanding of the SM: (a.1) $\Gamma (\Lambda_c^+ \to e^+\nu X_s)$ $\sim $ $\Gamma (D^0 \to e^+\nu X_s)$; (a.2) $\Gamma (\Lambda _c^+)$ $\sim$ $2 \; \Gamma (D^0)$. 
Of course, this is still within one sigma, indicating that additional data \& complex analyses are required. 

\noindent 
(b) Large differences exist between the lifetimes of the charm baryons, specifically, a factor of
$\sim$ 4 between $\tau (\Xi_c^+)$ \& $\tau (\Xi_c^0)$, with $\tau (\Lambda_c^+)$ being in the middle. 
Furthermore, $\tau (\Xi_c^+)/\tau (\Omega_c^0) \sim 6.4$. We know that differences of ${\cal O}(1/m_c)^2$ have already appeared, but not for charm mesons in particular. However it has been determined that
${\cal O}(1/m_c)^2 \sim {\cal O}(1/m_c)^3$ is numerical. Of course, we hope to measure these lifetimes
more accurately. 

\noindent 
(c) There is a good motivation for considering how to overcome difficult challenges in order to measure the
semi-leptonic branching ratios of $\Xi_c^{0,+}$ and even $\Omega_c^0$.

In the future, LQCD will be tested on new \& important avenues of research.

\subsection{CP asymmetries in two-, three-, \& four-body FS}

The SM gives us basically zero weak phases on DCS transitions because of
$V^*_{cd}V_{us}$ in $c\to u \bar sd$, and very small weak phases in SCS transitions because of
$V^*_{cd}V_{ud}$, $V^*_{cs}V_{us}$, \&  their interferences.
However, the connections between the worlds of the hadrons and quarks (\& gluons) are complex.
We expect the impact of SM penguin diagrams in the latter. The question is: how much and where?
SM penguin diagrams are affected by the difference between $V^*_{cd}V_{ud}$ \& $V^*_{cs}V_{us}$, but primarily depend on long-distance dynamics in charm transitions. This means that they produce re-scattering. Penguin diagrams indicate the direction to take to include FSI in SCS decays semi-quantitatively {\em at best}, but do not indicate any direction as regards DCS transitions.
Furthermore, the re-scattering amplitude landscape is significantly broader than penguin diagrams; it also produces DCS amplitudes. ND have a significantly greater effect on CP asymmetries than on rates, because of interferences $\propto T^*_{\rm SM} T_{\rm ND}$.

To date, neither direct nor indirect CP asymmetries have been found in charm hadrons, where we have probed
SCS transitions. The obtained data are closing in on a case where one might expect CPV in the SM. Asymmetries in DCS transitions have not been probed.

Non-leptonic amplitudes of charm hadrons are primarily given by two-, three-, \& four-body FS
\footnote{In Refs.\cite{CICERONE,IBCHINA}, I discussed true FS such as
$D \to \pi \pi$, but also pseudo-two-body $D \to \pi \rho$ \&  $\rho \rho$.}.
CP asymmetries in true two-body FS give numbers ``only", while Dalitz plots provide
two-dimensional asymmetries and significantly more for four-body FS. Obviously, one
first focuses on two-body FS for both experimental \& theoretical reasons, and our
research community generally adopted this approach in the past. Golden \& Grinstein \cite{GOLDEN} were the first to discuss
three-body FS in $D$ decays using non-trivial theoretical tools. Novel and refined tools have
appeared, which I will discuss below.

Measuring averaged asymmetries is only the first step. Accurately probing regional
asymmetries is crucial.
This shows that FSI, including broad resonances such as $\sigma/f_0(500)$ \&
$\kappa/K^*_0(800)$ \footnote{The latter has not yet been established.}
in the world of hadrons,
change the landscape of the world of quarks significantly. However, predicting this quantitatively poses a veritable challenge.

We must probe data in ``model insensitive" ways. There are several roads towards obtaining the necessary
information and some examples can be seen in \cite{REIS,WILL}. Comparing the results of these studies reveals both their strong and weak points. However, these {\em cannot} be the
final steps.
The real underlying dynamics do not always provide the best data fits, as evidenced in the long history of this field. We require further thought, redefined tools and, in particular, consideration of correlations with other FS based on CPT invariance.
In the future, dispersion relations \cite{DR,KUBIS} based on low-energy collisions of
two hadrons will be used; their power lies in combining data with
experimental \& theoretical tools.

Then, one must also probe regional CP asymmetries using different technologies.
The ratios of regional asymmetries do not depend on production asymmetries. One obtains more observables to check experimental uncertainties. On the theory side, considerably more work is required, but this allows theoretical uncertainties regarding the impact of non-perturbative QCD and the impact of the existence of ND and its features to be examined. We have seen that {\bf FSI} have a large impact, on suppressed decays of charm
(\& beauty) hadrons in particular. Probing
three- \& four-body FS provides  us with an indication of the sizable amount of work required on both the experimental \& theoretical sides; however, there will be ``prizes" at least for a deeper understanding of strong forces, and even more in the future concerning ND. I will not give a complete review; instead, I will focus on a few cases only.

With many-body FS,
one can describe SCS amplitudes by adding pairs of $\bar qq$ to
$c \to u \bar s s$ and $c \to u \bar d d$ and penguin $c \to u \bar q q$ with $q = u,d,s$.
Drawing and looking at diagrams is one approach; however, to
calculate their amplitudes even semi-quantitatively is another issue entirely. This is due to
non-perturbative QCD. The re-scattering strength depends on $\bar q_i q_i \to \bar q_jq_j$, where $i$ describes flavor rather than color. Likewise, for DCS decays for $c\to d u \bar s$ without penguin diagrams, one ``simply" adds pairs of $\bar q q$. {\em Regional} CP asymmetries do not depend on production rates; furthermore, they provide us with additional information about the underlying dynamics.

\subsubsection{CPV in $D^0$ transitions}
\label{CPVMES}

One can examine diagrams of
SCS $D^0$ decays in the world of quarks: $[c \bar u] \to u \bar d d \bar u$,
$[c \bar u] \to u \bar s s \bar u$, and penguin $[c \bar u] \to u \bar q q \bar u$ with $q = u,d,s$.
There are three important points to note:

(1) When one considers these diagrams for two-body FS $[\bar q_i q_j]$, one should also measure
$D^0 \to 2\pi^0$, $\pi^0 \eta$, $\pi^0  \eta^{\prime}$, etc. This one is obvious;, however, the following
two points are more subtle;

(2) Very differently from kaon decays, SCS FS of $D^0$ two-body non-leptonic decays do not dominate. Three- \& four-body FS constitute the majority of the process. It is easy to add one or two pairs of $\bar qq$ to the
diagrams. However, this does not mean that we have true control over the FS landscape, and we require the assistance of
more theoretical tools. In particular, we cannot focus on two-body FS and
probe with U-spin symmetry;

(3) Again, it is easy to draw diagrams; however, this does not mean we understand the dynamics. The cases of $B^0$ and $D^0$ decays and the impact of penguin diagrams differ significantly.

The landscape is more complex for $D^0$ than $D^+_{(s)}$, since indirect CPV affects
$D^0$ transitions only. It depends on $D^0 - \bar D^0$ oscillations and, therefore, on their times.
So far, we have primarily focused on two-body FS. For
SCS, $D^0 \to K^+K^-$, $\pi^+\pi^-$ (\&, in the future, $D^0 \to 2\pi^0$ and
$\bar K^0 K^0$, $\pi^0\eta$) and, for DCS, $D^0 \to K^+\pi^-$, where one can measure
$x_D = \Delta M_D/\Gamma _{D^0}$ \& $y_D= \Delta \Gamma_D/2\Gamma_D$ in
different transitions. There are four points to consider:

(a) CP asymmetries have two sources here, i.e., $D^0-\bar D^0$ oscillations and FS;

(b) The SM can minimally produce direct CPV in DCS decays;

(c) ND could have an impact on both, i.e., indirect CPV
\footnote{ND may have an impact on $x_{D^0}$.} and direct CPV that depend on the FS;

(d) It is crucial to measure the correlations between these sources with accuracy.

From the theoretical perspective, it is best to probe CP asymmetries in DCS, i.e.,
$D^0 \to K^+\pi^-$ vs. $\bar D^0 \to K^- \pi^+$, using large data, where
\bea
\frac{{\rm rate}(D^0(t) \to K^+\pi^-)}{{\rm rate}(\bar D^0 (t) \to K^-\pi^+)} &=&
\frac{|T(D^0 \to K^+\pi^-)|^2}{|T(\bar D^0 \to K^-\pi^+)|^2} \cdot \\
&& \cdot [1 +Y_{K\pi}(t\Gamma_D) +Z_{K\pi}(t\Gamma_D)^2]  \; ,\\
Y_{K\pi} &=& \frac{y_D}{\tan ^2 \theta_C}{\rm Re} \left(\frac{q}{p}\hat \rho _{K\pi} \right) +
\frac{x_D}{\tan ^2 \theta_C}{\rm Im} \left(\frac{q}{p}\hat \rho _{K\pi} \right), \\
Z_{K\pi} &=& \left| \frac{q}{p} \right|^2 \cdot \frac{x_D^2 +y_D^2}{4\tan ^2 \theta_C}
\cdot |\hat \rho _{K\pi}  |^2,
\eea
which can be written as
\beq
\frac{q}{p}\frac{T(\bar D^0 \to K^+\pi^-)}{T(D^0 \to K^+\pi^-)}
= e^{i\delta_{K\pi}} \frac{1}{{\rm tan}^2 \theta_C} \hat \rho _{K\pi} \; .
\eeq
This leads to the strong phase $\delta_{K\pi}$ depending on the FS with
$x_D^2 +y_D^2 = (x_D^{\prime})^2+(y_D^{\prime})^2$, where
\beq
y^{\prime}_D = y_D \, {\rm cos} \delta_{K\pi} - x_D \, {\rm sin}\delta_{K\pi} \: , \;
x^{\prime}_D = x_D\,  {\rm cos} \delta_{K\pi} + y_D\, {\rm sin}\delta_{K\pi} \; .
\eeq
The ``natural scale" for DCS is $\tan ^2 \theta_C$ with $|\hat \rho _{K\pi} | \sim {\cal O}(1)$. This also means that the natural scale for indirect
CPV is significantly enhanced by $(x_D,y_D)/\tan ^2 \theta_C$. Again, reward vs. cost must be considered.

SCS decays give BR$(D^0 \to K^+K^-) \simeq 4 \cdot 10^{-3}$ and
BR$(D^0 \to \pi^+\pi^-) \simeq 1.4 \cdot 10^{-3}$ on their ratios. This indicates the large impact of
re-scattering.
The FS of $2\pi$ produces $I = 0,2$, while $\bar KK$ with $I=0,1$; thus, re-scattering occurs for $I=0$ FS because of strong forces. $D^0$ transitions also produce neutral hadrons in the FS, i.e.,
$\bar K^0 \pi^0$, $\bar K^0K^0$, $2\pi^0$, \& $K^0 \pi^0$ (even ignoring the $\eta$ meson).
However, we have little quantitative control (at present).

Obviously, one uses CPT invariance. One applies
G parity that connects two, four, \& six pions and three \& five pions.
It has been suggested \cite{GRONAU} that U-spin symmetry should be probed using probe amplitudes with 
$\frac{\sqrt{|A(D^0 \to K^+K^-)A(D^0 \to \pi^+\pi^-)|}}{\sqrt{|A(D^0 \to K^+\pi^-)A(D^0 \to K^-\pi^+)|}}$
 = 1. This is to compare two SCS amplitudes against
one Cabibbo-favored \& DCS amplitudes in two-body FS for unity. I disagree strongly with this ``tool" as regards understanding the
underlying dynamics: first, we should not focus on charged pions \& kaons and, second,
many-body FS should not be ignored. The landscape is complex.

To further clarify this point, I ignore $D^0 - \bar D^0$ oscillations
\footnote{$D^0 - \bar D^0$ oscillations also include interference between Cabibbo-favored
\& DCS amplitudes.}. One examines (refined) tree diagrams
$[c\bar u] \to u\bar d d \bar u$ and $[c\bar u] \to u\bar s s \bar u$.
For the parametrizations shown here, only the second produces a weak phase
from $V_{cs}$. However, one requires correlations between both because of the strong FSI. A penguin
{\em diagram} shows $ [c\bar u] \Longrightarrow u \bar u$,
where one adds pairs of $\bar q_i q_i$ and $i$ indicates light-flavor quarks, $u,d,s$.
Re-scattering $q_i \bar q_i \to q_j \bar q_j$ with $i \neq j$ is also used to produce
$[c\bar u] \to u\bar q_i q_i \bar u$ in principle, but no more. However, we can continue to obtain
$[c \bar u] \to u \bar d q_i\bar q_i d \bar u$, $u\bar s q_i \bar q_i s \bar u$,
$u\bar q_i q_i \bar q_j q_j \bar u$, etc., to produce $D^0 \to \pi^+\pi^-\pi^0$, $3\pi^0$, $K^+K^-\pi^0$,
$\bar K^0 K^0 \pi^0$, etc., and $D^0 \to 2\pi^+2\pi^-$, $K^+K^- \pi^+\pi^-$, $4 \pi^0$, etc.
This occurs the majority of the time. Therefore, three- \& four-body FS have considerable impact on non-leptonic decays.

Ref.\cite{3PIONS} for $D^0 \to \pi^+\pi^- \pi^0$ seems to ignore
re-scattering from $[c\bar u] \to u\bar s s \bar u$; I see no reason at all to justify this. This is another example of why we must consider real re-scattering rather than simply looking at diagrams.
We know that strong phases are large in general and depend
on the FS. We must connect amplitudes in the world of quarks with those of hadrons using theoretical
tools \& judgment.

Effective transition amplitudes with re-scattering connect two-body FS of charged \& neutral hadrons (see Sect.(\ref{EFFECT})), i.e., U- \& V-spin symmetries are affected in the world of hadrons.
If the U-spin violations are quite small and, therefore, the expansion
of U-spin violations makes sense, I would regard this as ``luck",
or else we have overlooked some important features of non-perturbative QCD.

For DCS decays, one describes $c \to d \bar s u$ amplitude $V^*_{cd}V_{us}$ with basically
zero weak phases, see Eq.(\ref{MATRIX}). This approach is excellent
for finding ND and, perhaps, its features also.
Penguin diagrams do {\em not} help to describe re-scattering/{\bf FSI} here.

One can measure $D^0 \to K_S \pi^+\pi^+$ and probe its Dalitz plot including interference
between Cabibbo-favored \& DCS amplitudes. This was performed for the
$K_S\rho^0$, $K^{*,-}\pi^+$ \& $K^{*,+}\pi^-$ resonances, which are somewhat narrow. The Particle Data Group (PDG) provides
data about averaged
CP asymmetries only, with $A_{\rm CP} (D^0 \to K_S \pi^+\pi^-) = (-0.1 \pm 0.8)\%$ \cite{PDG14}. We require additional data and more refined analyses incorporating broad resonances  with {\em regional} asymmetries. These should be compared with $A_{\rm CP} (D^0 \to K_S K^+K^-)$. Likewise, we must apply this approach to
$D^0 \to 3\pi$, $\pi \bar KK$. The next steps are to probe regional CP asymmetries in
DCS $D^0 \to K^+\pi^-\pi^0$ and $D^0 \to 4\pi$, $\bar K K \pi \pi$.

Hadronic uncertainties in $c \to u$ decays have been discussed \cite{BEVMEAD}, in particular as regards $D^0 \to \pi^+\pi^-$, $\rho^+\rho^-$, $\rho \pi$
\footnote{It does not matter that the first results from
LHCb CP asymmetries have disappeared.}.
I disagree with some of the statements in this paper. The title $c \to u$ suggests that there is
no difficulty in connecting diagrams with operators; furthermore, the authors focus primarily on tree and
penguin diagrams in SCS transitions. There are subtle, but important differences between diagrams, local, \& non-local operators; we must overcome non-trivial challenges there. Again, the left sides of Eqs. (\ref{CPTAMP1},\ref{CPTAMP2}) describe hadron amplitudes, while the right sides incorporate bound states of $\bar q_i q_j$. It is crucial to measure true three- and
four-body FS with accuracy.

LHCb data has given us integrated refined T-odd measurements for
$D^0 \to K^+K^-\pi^+\pi^-$ \cite{LHCbODD}, where
\bea
A_T &= &(-7.18 \pm 0.41(\rm stat) \pm 0.13(\rm syst))\%,  \\
\bar A_T &=& (-7.55 \pm 0.41(\rm stat) \pm 0.12(\rm syst))\%,\\
a_{\rm CP}^{T-{\rm odd}} &=& (0.18 \pm 0.29(\rm stat) \pm 0.04(\rm syst))\%.
\eea
I wish to emphasize four points:

\noindent
(a) The values for $A_T$ \& $\bar A_T$ are not large, but they are still sizable because of re-scattering
at low energies;

\noindent
(b) Obviously averaged  CP asymmetry is consistent with very small SCS decays; however,
a semi-regional asymmetry could be sizable;

\noindent
(c) CPT invariance due to correlations of $D^0 \to K^0\bar K^0\pi \pi$,
$4 \pi$, etc., should not be forgotten;

\noindent
(d) Finally, one must probe CP symmetries in DCS $D^0 \to K^+\pi^+\pi^+\pi^0$, etc.

\subsubsection{CP asymmetries in $D^+$ \& $D^+_s$ decays}

The best locations to find CP asymmetries are $D^+ \to \pi^+\pi^+\pi^-$, $\pi^+K^+K^-$.
The PDG lists only averaged CP asymmetries for SCS \cite{PDG14}, with
\bea
A_{CP}(D^+ \to \pi^+\pi^+\pi^-) &=& (-2 \pm 4) \, \%, \\
A_{CP}(D^+ \to \pi^+K^+K^-) &=& (0.36 \pm 0.29) \, \%.
\eea
It is very important to probe regional and similar asymmetries for $D^+ \to \pi^+\pi^+ \pi^-\pi^0$, $K^+K^- \pi^+ \pi^0$, etc.

Even averaged CP asymmetries in DCS have not been measured to date. In the future, it will be crucial (but not easy) to probe regional asymmetries in $D^+ \to K^+\pi^+\pi^-$, $K^+K^+K^-$, \& their correlations. There is
almost no background from the SM and, therefore, this provides a golden opportunity for establishing ND. This also applies to $D^+ \to K^+\pi^+\pi^-\pi^0$.

For SCS rates, the data provides us with information concerning the asymmetries \cite{PDG14}, where
\bea
A_{CP} (D^+_s \to K_S\pi^+) &=& (1.2 \pm 1.0) \%, \\
\langle A_{CP}(D^+_s \to K^+\pi^+\pi^-)\rangle &=& (4\pm 5 )\, \%.
\eea
Of course, we do not expect any non-zero values on that level.
Obviously, additional data and probing of regional asymmetries are required in the future.
Furthermore, we require considerably more data concerning DCS asymmetries, in particular for
$D^+_s \to K^+K^+\pi^- $.
This will provide us with a more ``exotic" landscape for CP asymmetries.

\subsubsection{CP asymmetries in non-leptonic charm baryon decays}
\label{CPVBAR}

Refined tree \& penguin diagrams
and re-scattering are important, as in the case of mesons, but WA diagrams are not
suppressed by chiral symmetry for baryons. Therefore, additional operators must be included.

One example for DCS is that one can compare $\Lambda_c^+ \to p K^+\pi^-$ and
$\bar \Lambda_c^- \to \bar p K^-\pi^+$; this can be calibrated through Cabibbo-favored decays,
$\Lambda_c^+ \to p K^-\pi^+$ and
$\bar \Lambda_c^- \to \bar p K^+\pi^-$. Furthermore, for SCS, we can compare $\Lambda_c^+ \to p \pi^+\pi^-/p K^+K^-$ and
$\bar \Lambda_c^- \to \bar p \pi^+\pi^-/\bar p K^+K^-$, which can be calibrated as indicated above.
The landscape is even richer with $\Xi_c^{0,+}$ \& $\Omega_c^0$.

\section{Dynamics of $\tau$ leptons}
\label{TAUDYN}

One can probe $\tau$ leptons in relation to flavor violation, e.g.,
$\tau ^- \to l^- \gamma$, $l^-\mu^-\mu^+$, $l e^+e^-$, $l^- h$, with $l = e, \mu$
\& $h= \pi$, $K$, $\pi \pi$, $K \bar K$, etc. \cite{PASS2,PASSCH,YOGI}.
\begin{figure}
\begin{center}
\includegraphics[width=1.\textwidth]{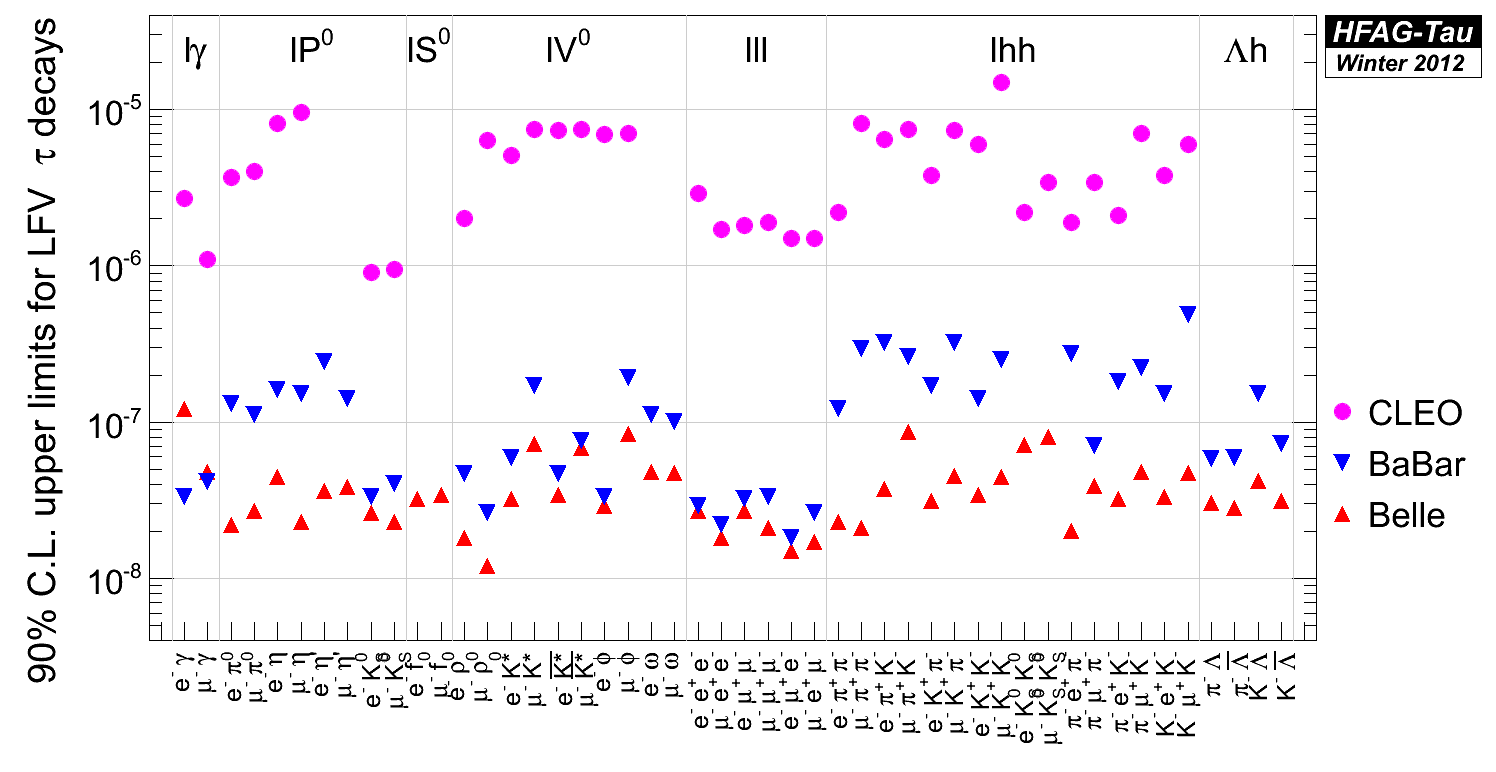}
\caption{\it Bounds on Tau Lepton Flavor Branching Ratios from CLEO, BaBar, Belle. Figure taken from HFAG~\cite{hfag-2014}}
\label{fig:Tau_LFV_HFAG}
\end{center}
\end{figure}
The crucial challenge is that we require extremely large numbers of $\tau$ decays, although we can control the SM background before we obtain candidates in a complex landscape. When candidates are found, disagreements regarding rates
($\propto |A_{\rm ND}|^2$) will begin between theorists ``in the field". However, we have the tools to deal with
this in a detailed manner \cite{PASS2,PASSCH}. In fact, this would be a ``field day" for the theorists.
However, one should not forget the differences between ``statistical" and ``systematic" uncertainties;
the latter are not easily accepted when additional data are available.

The most obvious goal is to extract $|V_{ud}|$ \& $|V_{us}|$ from semi-hadronic $\tau$ decays and, through comparison, identify which yields semi-leptonic $D$ decays. However, I discuss CP asymmetries here, which are subtle and depend
on interference between SM \& ND amplitudes.

\subsection{$\tau$ decays}

CPT invariance predicts
\bea
\Gamma (\tau ^- \to \nu X^-_{S=0}) & = & \Gamma (\tau ^+ \to \bar \nu X^+_{S=0}), \\
\Gamma (\tau ^- \to \nu X^-_{S=-1}) & = & \Gamma (\tau ^+ \to \bar \nu X^+_{S=+1}).
\eea
Measuring $X_{S=0}$ with accuracy tests our understanding of non-perturbative QCD forces.
It seems we have the optimum opportunity to find CPV in different
$X^-_{S=-1} = K^-$, $K^-\pi^0$, $K^0 \pi^-$, $K^-\pi^+\pi^-$, etc.
Present data concerning CPV in SCS decays
$\tau ^+ \to \nu K_S \pi^+ [+\pi ^{0\, \prime} {\rm s}]$ indicate a difference of 2.9-$\sigma$ between SM predictions due to the
well-known $K^0 - \bar K^0$ oscillation, where
\bea
A_{\rm CP}(\tau ^+ \to \bar \nu K_S\pi^+)|_{\rm SM} &=& +(0.36 \pm 0.01) \%, \; \; \; \cite{TAUBS}
\label{TAU1}
\\
A_{\rm CP}(\tau ^+ \to \bar \nu K_S\pi^+ [+\pi^{0\; \prime} {\rm s}])|_{\rm BaBar2012} &=&
-(0.36 \pm 0.23 \pm 0.11) \%
\; \; \cite{BABARTAU} \;.
\label{TAU2}
\eea
(Note the sign.) A 2.9-$\sigma$ difference is not a significant achievement, however,
one must measure CPV in $\tau^+ \to \nu K^+\pi^0 $,
$\nu K^+\pi^+\pi^-$, etc., and consider correlations due to CPT invariance.

The available data concerns measurements of integrated CP asymmetries only. We must probe
regional CP asymmetries, and we must wait for Belle II
(and the Super Tau-Charm Factory if \& when it is completed). If
polarized $e^+e^-$ beams exist \cite{TAUD+}, that would be wonderful.

One must accurately compare
regional data for $\tau ^+ \to \nu \pi^+\pi^0$,
$\nu \pi^+\eta$, $\nu \pi^+\pi^+\pi^-$, $\nu \pi^+\pi^0\pi^0$, etc.
This is a test of experimental uncertainties and identifying CPV in this case is quite difficult. These items
will soon be discussed in detail \cite{PASSCH}.

One must accurately measure the correlations with DCS
$D^+ \to K^+\pi^+\pi^-$/$K^+K^+K^-$, etc. In general, I would say that
these $\tau$ decays provide data that can be used to probe DCS \& SCS in $D$ decays that must be calibrated. Furthermore -- although unlikely -- it might reveal impact towards dark matter.

\subsection{Production of $\tau$ lepton pairs to probe their electric dipole moments}
\label{EMSTAU}

The difference between the SM predictions for $(g - 2)_{\mu}$ and the data on the 3-$\sigma$ level
attracted considerable interest in our community. This was also the case regarding the electric dipole moments (EDMs) of electrons, neutrons, etc. \cite{DEKENS}.
Furthermore, it has enhanced the interest about the $\tau$ EDM \cite{BERN1,PICH,PASS,PASSCH}.
Transitions of $\tau^+\tau^-$ between production and decays are subtle. We know the SM has no
connection with the huge asymmetry of matter and anti-matter. Thus, it is possible that the source of this
asymmetry may be found in EDMs, or in their connection with others. No EDMs for electrons, muons, atoms, nuclei, etc., have been found (yet).

A search was conducted for $d_{\tau}$ in
$e^+e^- \to \tau^+\tau^-$ measurements \cite{INAMI}, where
\beq
-0.22 <{\rm Re} (d_{\tau}) < 0.45 \; [10^{-16} e\cdot {\rm cm}] \; \; , \; \;
- 0.25 < {\rm Im} (d_{\tau}) < 0.08 \; [10^{-16} e\cdot {\rm cm}].
 \eeq
There are some subtle points to note. One can discuss the weak dipole moment through effective
$Z^0$ couplings \cite{PICH}. However, at present, certain limits apply \cite{ALEPH}
\beq
|{\rm Re} (d^W_{\tau})| < 0.5 \; [10^{-17} e\cdot {\rm cm}] \; \; , \; \;
|{\rm Im} (d^W_{\tau})| < 1.1 \; [10^{-17} e\cdot {\rm cm}].
\eeq
The probability of finding $\tau$ EDMs is small, but not zero. Therefore, this merits thorough investigation; if this does not appeal to you, you are in the wrong ``business".

\section{Connections with beauty hadron decays}
\label{BEAUT}

Direct CPV  $\sim 0.1$ was predicted \cite{1988BOOK,WOLFFSI}
in $\bar B_d \to K^-\pi^+$ 
in relation to the impact of strong re-scattering
(\& identified theoretical uncertainties); it was found at a significantly later date. 
{\bf FSI} produces not only a complex landscape for many-body FS,
but also has a large impact.

\subsection{CP asymmetries in $B^{\pm}$ with CPT invariance}
\label{CPVBEAUHAD}

The CKM-suppressed weak decays of beauty hadrons produce FS with more hadrons than
two-, three-, \& four-body FS.
Therefore, one expects that CPT invariance is not a ``practical" tool as regards beauty decays; however, surprising findings were obtained. Data indicate that CKM-suppressed $B$ decays primarily populate the boundaries of Dalitz plots, while
the centers are close to being empty. At the qualitative level, one should not be surprised.
CPV is caused by interference; therefore, one expects large regional asymmetries. However, the extent and location of these asymmetries is unclear.

\subsubsection{$B^{\pm} \to K^{\pm}\pi^+\pi^-$ vs. $B^{\pm} \to K^{\pm}K^+K^-$}

LHCb data show sizable CP asymmetries averaged over the FS \cite{LHCb028}, where
\bea
\Delta A_{CP}(B^{\pm} \to K^{\pm} \pi^+\pi^-) &=&
+0.032 \pm 0.008_{\rm stat} \pm 0.004_{\rm syst}
\pm 0.007_{\psi K^{\pm}},
\label{SUPP1}
\\
\Delta A_{CP}(B^{\pm} \to K^{\pm} K^+K^-) &=&
- 0.043 \pm 0.009_{\rm stat} \pm 0.003_{\rm syst}
\pm 0.007_{\psi K^{\pm}},
\label{SUPP2}
\eea
with 2.8-$\sigma$ \& 3.7-$\sigma$ from zero.
The sizes of these averaged asymmetries are not surprising; however, this does not mean that
they could be predicted. It is very
interesting that they have opposite sign as a result of CPT invariance.

LHCb data indicate regional CP asymmetries \cite{LHCb028}, where
\bea
A_{CP}(B^{\pm} \to K^{\pm} \pi^+\pi^-)|_{\rm regional} &=& + 0.678 \pm 0.078_{\rm stat}
\pm 0.032_{\rm syst}
\pm 0.007_{\psi K^{\pm}},
\label{SUPP3}
\\
A_{CP}(B^{\pm} \to K^{\pm} K^+K^-)|_{\rm regional} &=& - 0.226 \pm 0.020_{\rm stat}
\pm 0.004_{\rm syst} \pm 0.007_{\psi K^{\pm}} \; .
\label{SUPP4}
\eea
Regional CP asymmetries are defined by the LHCb collaboration. Positive asymmetry is at low
$m_{\pi ^+\pi ^-}$, directly below $m_{\rho^0}$, while negative asymmetry occurs at both low and high $m_{K^+K^-}$ values. The opposite signs in Eqs.(\ref{SUPP1},\ref{SUPP3}) and Eqs.(\ref{SUPP2},\ref{SUPP4}) should be noted. This is an effective approach, but (in my view) this is not the final step.
In the future, we must:
(a) analyze the data using insensitive techniques \cite{REIS,WILL}; and
(b) remember that the true underlying dynamics do not often give the best
fitting analyses. There is sizable room for thinking, i.e., superior theoretical tools should be used for strong {\bf FSI}, such as dispersion relations.

One expects large regional
CP asymmetries, but the locations and size are unclear. Thus, researchers focus on small regions in the Dalitz plots while the
centers are mostly empty. I am surprised by this; however, this approach does provide us with further highly non-trivial information concerning non-perturbative QCD.

\subsubsection{$B^{\pm} \to \pi^{\pm}\pi^+\pi^-$ vs. $B^{\pm} \to \pi^{\pm}K^+K^-$}

One expects smaller rates of these FS based on ``experience" of $\bar B_d \to K^-\pi^+$
vs. $\bar B_d \to \pi^+\pi^-$ cases. Indeed, this is true for the data given above.
However, comparing CP asymmetries
reveals the surprising impact of penguin diagrams \cite{IRINA}, as
\bea
\Delta A_{CP}(B^{\pm} \to \pi^{\pm}  \pi^+\pi^-) &=&
+0.117 \pm 0.021_{\rm stat} \pm 0.009_{\rm syst}
\pm 0.007_{\psi K^{\pm}},
\label{SUPP5}
\\
\Delta A_{CP}(B^{\pm} \to \pi^{\pm} K^+K^-) &=&
- 0.141 \pm 0.040_{\rm stat} \pm 0.018_{\rm syst}
\pm 0.007_{\psi K^{\pm}}  \;.
\label{SUPP6}
\eea
That is, one obtains information concerning the amplitudes $T(b \Rightarrow s)$ $\gg$ $T(b \Rightarrow d)$ in the SM.
Again CPV appears in small regions in the Dalitz plots, while the
centers are mostly empty \cite{IRINA}.
\bea
\Delta A_{CP}(B^{\pm} \to \pi^{\pm} \pi^+\pi^-)|_{\rm regional} &=&
+0.584 \pm 0.082_{\rm stat} \pm 0.027_{\rm syst}
\pm 0.007_{\psi K^{\pm}},
\label{SUPP7} \\
\Delta A_{CP}(B^{\pm} \to \pi^{\pm} K^+K^-)|_{\rm regional} &=&
- 0.648 \pm 0.070_{\rm stat} \pm 0.013_{\rm syst}
\pm 0.007_{\psi K^{\pm}}  .
\label{SUPP8}
\eea
As previously, averaged CP asymmetries require large asymmetries in small regional areas.
One should note not only the strengths of these asymmetries, but also their signs.
The impact of
broad scalar resonances vs. narrow resonances (such as those due to dispersion relations) should also be discussed.

\subsection{CP asymmetries in beauty baryons}
\label{CPBBAR}

To date, CP asymmetries have been probed in two-body FS \cite{PDG14}, where
\bea
A_{\rm CP}(\Lambda^0_b \to p \pi^-) &=& 0.03 \pm 0.18, \\
A_{\rm CP}(\Lambda^0_b \to p K^-) &=& 0.37 \pm 0.17.
\eea
It seems to me that one ``promising" channel in three-body FS remains, i.e.,
$\Lambda^0_b \to \Lambda D^-\pi^+$, where we can probe for regional
asymmetries without production asymmetry. In the future one could measure
$\Xi_b^0 \to \Lambda \pi^+\pi^-/\Lambda K^+K^-$,
$\Xi_b^- \to \Lambda \pi^-\pi^+\pi^-$, $\Lambda \pi^-K^+K^-$, etc.

\section{Summary \& outlook for the future}
\label{FUT}

\begin{figure}[h!]
\begin{center}
\includegraphics[width=10cm]{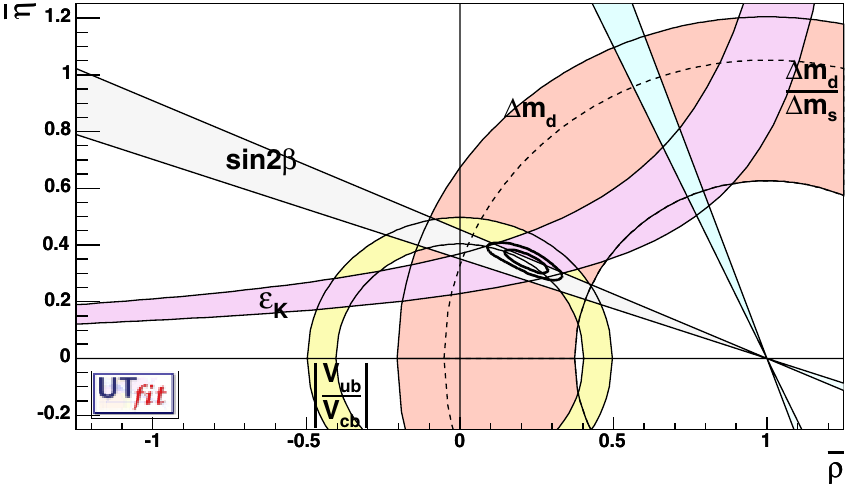}
\end{center}
\caption{Correlations between other triangles}
\label{fig:TRIANGLES}
\end{figure}

As I have said in the introduction: my goal was to ``paint" a picture of the landscape of heavy-flavor
hadrons and charged leptons and discuss correlations between them.
Furthermore, charm hadrons
and $\tau$ leptons act as gateways from light to heavy flavors.

First, I described this qualitatively:
\begin{itemize}
\item
We have an even richer experimental landscape in charm \& beauty hadrons and $\tau$ leptons for the future,
as regards existing experiments at the LHCb (\& perhaps the A Toroidal LHC Apparatus (ATLAS) \& the Compact Muon Solenoid (CMS)),  Beijing Spectrometer (BES) III, and Belle II. Their programs
will deepen our understanding of fundamental dynamics.
Plans are also in place for construction of the Super Tau-Charm Factory, Super-$Z^0$ Factory, etc.;

\item
It is not sufficient to measure two-body FS, as we must also probe three- \& four-body FS.
We must transition from the accuracy era to a precision-based period using the optimum tools (such as
dispersion relations) that require a connection between theory \& experiment;


\item
Searching for a golden medal is {\em not} enough. It is
crucial to consider and measure correlations in flavor transitions;

\item
Using averaged strong phases is the first step. However, it is obvious that we must go beyond that
and measure regional phases. The ``road" towards three-body FS is obvious (in the world of
theorists). For four-body FS, however, the landscape is more ``complex". First, we can probe averaged
asymmetries, then moments (\& correlations with different definitions), and then
semi-regional asymmetries. Of course, appropriate thought should be applied.
\end{itemize}

Now I summarize the main points of this article semi-quantitatively:

(a) Flavor dynamics in the SM with three families of up- \& down-quarks are described
by six triangles, as shown in detail above. It is crucial to probe their correlations.
There is a well-known example, namely, the ``golden" triangle for $B$ transitions, with limits
given by $\Delta M_{B_d}/\Delta M_{B_s}$ \& $\epsilon_K$ (see Fig.\ref{fig:TRIANGLES});

(b) While QCD is the only local QFT that can describe strong forces, it is crucial to understanding non-perturbative
dynamics. We learn significantly more about its impact and its interaction with other tools, namely,
{\bf OPE}, {\bf HQE}, \& chiral symmetry. While charm hadrons primarily act as heavy-flavor particles, QCD tells us how they approach the limits of heavy-flavor states. In particular, charm transitions provide an excellent testing ground
for our quantitative control of LQCD.

(c) QCD shows the differences between {\em spectroscopy} and {\em weak dynamics}, but also their connections, which are often subtle. Charm hadrons, including baryons, and $\tau$ particles provide very good testing grounds.

(d) We must transition from {\em accuracy} to {\em precision} as regards fundamental dynamics. We require not only additional
data, but also additional \& superior theoretical tools. Therefore, I have given an overview of these tools
in Sect.\ref{TOOLS}.

(e) Obviously, I support the probing of CP asymmetries. However, this is for good reasons.
We probe interference between SM \& ND amplitudes. More importantly, we must measure
{\em regional} CP asymmetries with accuracy at the minimum \cite{PISA,RPP}.
If a researcher does not like difficult challenges, she/he is in the wrong ``business".

Table 1 compares the oscillation parameter landscape for neutral mesons, which is already
qualitatively rich. This landscape describes the transitions of the somewhat light-flavor meson $K^0$,
heavy $B^0$ \& $B^0_s$, and somewhat heavy $D^0$.
\begin{center}
\begin{tabular}{c|c|c|c}
\hline
$K^0$ & $D^0$ & $B^0$ & $B^0_s$ \\
\hline
$\Delta M_K \simeq \bar \Gamma_K$ & $\Delta M_D \ll \bar \Gamma_D$ &
$\Delta M_{B^0} \sim \bar \Gamma_{B^0}$ & $\Delta M_{B^0_s} \gg \bar \Gamma_{B^0_s}$ \\
$\Delta \Gamma _K \simeq 2 \bar \Gamma_K$ & $\Delta \Gamma _D \ll  \bar \Gamma_D$ &
$\Delta \Gamma _{B^0} \ll  \bar \Gamma_{B^0}$ &
$\Delta \Gamma _{B^0_s} \sim {\cal O}(\bar \Gamma_{B^0_s})$ \\
$\Delta \Gamma _K \sim \Delta M_K$ & $\Delta \Gamma _D \sim \Delta M_D$ &
$\Delta \Gamma _{B^0} \ll \Delta M_{B^0}$ & $\Delta \Gamma _{B^0_s} \ll \Delta M_{B^0_s}$ \\
\hline
\end{tabular}
\end{center}
As discussed in a long and detailed article \cite{RPP}, one can examine the plots of the mass distributions of $K_S$ vs. $K_L$
for $B_{d,L}$ vs. $B_{d,H}$ and $B_{s,L}$ vs. $B_{s,H}$ (on p. 1890--1891). The probing
of quark flavor dynamics will remain important in the future, but the landscape has changed. In particular,
we must transition from accuracy to precision, while also emphasizing correlations \cite{VARENNA}
\footnote{Computers can solve many problems, but not all; we should not forget {\em thought}.
One recent example: The autopilot on a plane in flight instructed the plane to
reduce altitude for no justifiable reason. Ultimately, the passengers and crew were saved because the experienced pilot succeeding in removing control of the plane from the computer. Is there an allegory in
this event?}.

\begin{figure}[h!]
\begin{center}
\includegraphics[width=6cm]{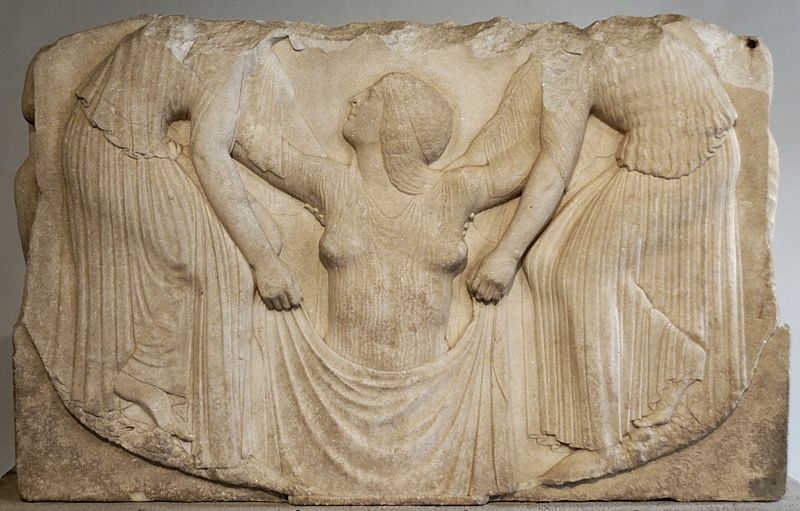}
\end{center}
\caption{Ludovisi throne: Goddess can be born with symmetry}
\label{fig:VENUS}
\end{figure}
Classical Greek art shows the connection between beauty and symmetry; an excellent example from Rome can be seen in Fig.\ref{fig:VENUS}.

\begin{figure}[h!]
\begin{center}
\includegraphics[width=4cm]{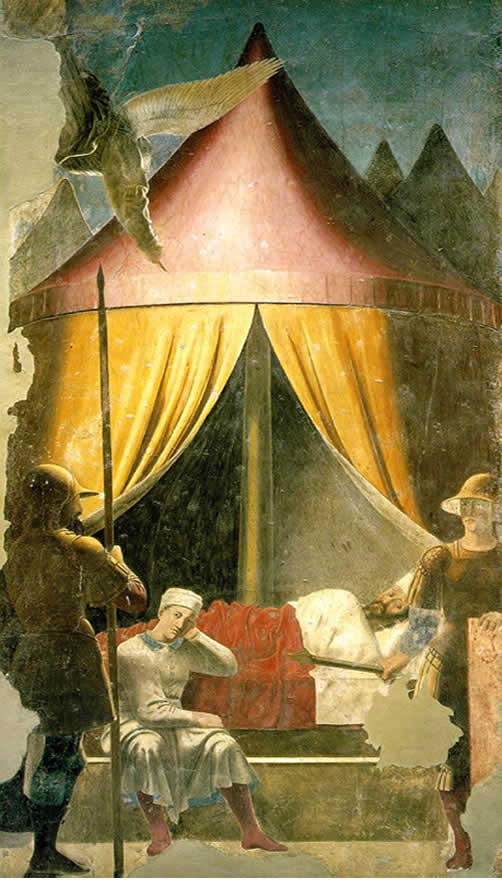}
\end{center}
\caption{Dreams}
\label{fig:PIERO}
\end{figure}

In the early Renaissance, approximately 1455 A.D., Piero della Francesa painted Constantine's Dream, depicting the dream of Constantine (the Great) the night before his crucial battle on the fringes of Rome in 312 A.D. Here, the connection between different dimensions can be seen (see Fig.\ref{fig:PIERO}).
Piero della Francesa was also known as a mathematician \& a geometer.

It is wonderful to dream about SUSY, etc., but we must be protected by
data. However, in the long-term, surprising results can be obtained. Allow me to finish with a personal comment:
for me, there should be a deeper, but subtle connection between fundamental dynamics and symmetry.
The data are ultimately the judges, but research timescales can be long, as evidenced by our experience.
I do not like to relinquish ``wonderful" ideas at an early stage.

{\bf Acknowledgments:} This work was supported by the NSF under grant number PHY-1215979. 
I am grateful to the Mainz Institute for Theoretical Physics ({\bf MITP}) for its hospitality and its partial support during the completion of this work. 

\vspace{4mm}


\end{document}